\documentclass[11pt]{article}
\usepackage{xcolor}
\usepackage{multirow}
\usepackage{enumerate}
\usepackage[scale=0.8]{geometry}

\usepackage{graphicx}
\usepackage[subrefformat=parens,labelformat=parens]{subfig}
\usepackage{hyperref}

\usepackage{subcaption}
\newcommand{\ds}{\displaystyle}

\usepackage{amsfonts,amsmath,amssymb}
\usepackage{multirow,multicol}
\usepackage{bbm}
\usepackage{xspace}
\usepackage{slashed}

\usepackage[sort&compress, english]{cleveref}
\newcommand{\K}{K\"ahler }

\usepackage[style=numeric-comp,url=true,doi=true,maxbibnames=99,sorting=none]{biblatex}

\bibliography{sample}

\numberwithin{equation}{section}

\usepackage{setspace}
\begin{document}
\phantom{}
\vspace{-.2cm}
\begin{center}
{

\LARGE {\bf Calabi--Yau Metrics with K\"ahler Moduli Dependence}\\[12pt]

\vspace{1cm}
\normalsize
{\bf{Andrei Constantin$^{a,b,}$}\footnote{a.constantin@bham.ac.uk}}
{\bf ,}
{\bf{Andre Lukas$^{b,}$}\footnote{andre.lukas@physics.ox.ac.uk}}
{\bf ,}
{\bf{Luca A. Nutricati$^{b,}$}\footnote{luca.nutricati@physics.ox.ac.uk}}
\bigskip}\\[0pt]
\vspace{0.23cm}
${}^a$ {\it 
School of Mathematics, University of Birmingham\\ Watson Building, Edgbaston, Birmingham B15 2TT, UK
}\\[2ex]
${}^b$ {\it 
Rudolf Peierls Centre for Theoretical Physics, University of Oxford\\
Parks Road, Oxford OX1 3PU, UK
}\\[2ex]
\end{center}
\vspace{15mm}

\begin{abstract}\noindent
We present a method to construct approximate analytic expressions for Ricci-flat K\"ahler metrics on Calabi--Yau threefolds with explicit dependence on the K\"ahler moduli. 
Our strategy combines numerical data obtained from machine learning with an explicit analytic Ansatz for the K\"ahler potential and symbolic regression methods. Specifically, we use neural networks to learn the K\"ahler potential at selected points in K\"ahler moduli space, fit this data to analytic expressions with K\"ahler moduli-dependent parameters, and determine an analytic form of these coefficients as functions of the K\"ahler moduli using symbolic regression. In this way, we reconstruct closed-form approximations to the Ricci-flat metric that retain explicit K\"ahler-moduli dependence.
We apply this method to two Calabi--Yau threefolds with $h^{1,1}=2$, namely a bicubic hypersurface in $\mathbb{P}^2 \times \mathbb{P}^2$ and a bi-degree $(2,4)$ hypersurface in $\mathbb{P}^1 \times \mathbb{P}^3$, both of which admit nontrivial discrete symmetry groups that simplify the structure of the metric.
In both cases, the resulting analytic expressions reproduce the numerically learned K\"ahler potentials with percent-level accuracy and yield a Ricci-flatness measure that remains sufficiently small across the sampled region. Our results represent a concrete bridge between purely numerical results for Calabi--Yau metrics and analytic constructions, opening the door to a systematic study of their dependence on K\"ahler moduli.

\end{abstract}

\setcounter{footnote}{0}
\setcounter{tocdepth}{2}
\clearpage

\section{Introduction}\label{sec:introduction}

A central objective of string phenomenology is to derive the parameters of the Standard Model from the geometry of extra dimensions. Recent advances suggest that this objective is no longer purely aspirational. With the development of efficient numerical algorithms and, in particular, neural-network methods for approximating Ricci-flat metrics, it has become possible to compute quantities in string compactifications, such as Yukawa couplings, fermion masses, and mixing angles, that are otherwise treated as input parameters in the Standard Model~\cite{Constantin:2024yxh,Butbaia:2024tje,Berglund:2024uqv}. Moreover, recent analyses indicate that the moduli space of Calabi--Yau (CY) compactifications may contain loci where the full set of Standard Model parameters can be simultaneously realised~\cite{Constantin:2025vyt}. Identifying and studying such regions requires detailed control over how physical observables vary across moduli space.

This dependence is highly nontrivial. While holomorphic quantities depend, at the perturbative level, only on complex structure, physical Yukawa couplings and kinetic terms require knowledge of the full Ricci-flat metric and its variation with the K\"ahler moduli. Numerical approaches have achieved impressive accuracy in constructing approximate
Ricci-flat metrics at given points in moduli space. Donaldson’s algorithm provides a systematic and provably convergent approximation scheme~\cite{Donaldson:2005hvr}, though achieving high accuracy can be computationally expensive. Moreover, since the algorithm produces metrics in integral K\"ahler classes proportional to the first Chern class of a chosen line bundle, varying the K\"ahler parameters requires changing the line bundle and, hence, the entire Ansatz. This makes it difficult to parametrise the metric continuously as a function of the K\"ahler moduli.

Over the years, a variety of numerical approaches to Ricci-flat metric construction have been developed~\cite{Headrick:2005ch, Donaldson:2005hvr, Douglas:2006rr, Braun:2007sn, Braun:2008jp, Headrick:2009jz,Cui:2019uhy}. More recently, machine-learning methods have enabled highly accurate and computationally efficient metric reconstructions~\cite{Ashmore:2019wzb, Anderson:2020hux,Jejjala:2020wcc,Ashmore:2021ohf,Larfors:2021pbb,Larfors:2022nep,Gerdes:2022nzr}. However, the resulting metric is represented implicitly by a trained neural network at fixed values of the moduli. Although this constitutes a functional representation, it does not provide a transparent analytic expression, nor does it make the moduli dependence explicit. As a consequence, extrapolating across moduli space or systematically analysing moduli dependence remains difficult.

Obtaining analytic expressions for the metric (even approximate ones) is therefore of central importance. They render the dependence on the moduli explicit, streamline computations, and allow one to track systematically how physical quantities vary across moduli space. Moreover, a proper understanding of moduli stabilisation requires that all relevant terms in the effective action be expressed explicitly in terms of the moduli, so that extrema can be identified and analysed in a controlled manner. Despite their importance, explicit analytic Ricci-flat metrics are largely unavailable for compact Calabi--Yau threefolds. While remarkable constructions exist for K3 surfaces, where hyperk\"ahler geometry and instanton methods have led to analytic descriptions of Ricci-flat metrics~\cite{Kachru:2018van, Kachru:2020tat}, no comparable analytic description is currently known for compact Calabi--Yau threefolds. 

Hybrid strategies combining numerical data with analytic reconstruction have begun to appear~\cite{Mirjanic:2024gek,Lee:2025pue}, 
indicating that such a synthesis is feasible. In this work, we develop such a hybrid approach for the K\"ahler-moduli sector. Concretely, we use neural networks to learn the Ricci-flat K\"ahler potential at selected points in K\"ahler moduli space, and then fit this data with an explicit analytic Ansatz whose coefficients are promoted to functions of the K\"ahler parameters. We then use symbolic regression on the fit data to find analytical expressions for these coefficients. In this way, we obtain closed-form approximate formulae for the metric that retain explicit and controllable dependence on the K\"ahler moduli. We restrict to Calabi--Yau threefolds with $h^{1,1}(X)=2$, in particular we study a bicubic hypersurface in $\mathbb{P}^2 \times \mathbb{P}^2$ and a bi-degree $(2,4)$ hypersurface in $\mathbb{P}^1 \times \mathbb{P}^3$. As we shall see, the resulting analytic expressions reproduce the numerically learned K\"ahler potentials to within a few percent, while making the moduli dependence fully explicit.
The associated CY metric, although not achieving the same accuracy as the neural network, remains Ricci-flat to a good approximation.

The paper is organised as follows. In Section~\ref{sec:strategy}, we describe the general strategy we adopt. We then proceed by applying this approach to the two Calabi--Yau manifolds mentioned above, presenting the results in Sections~\ref{sec:bicubic} and~\ref{sec:2_4}. We conclude in Section~\ref{sec:conclusions}.

\section{General strategy}
\label{sec:strategy}

\subsection{Motivation}

We begin with a brief review of Donaldson’s algorithm, which provides a systematic analytic approximation scheme for Ricci-flat metrics, but is structurally ill-suited to incorporate explicit dependence on the \K moduli.
Given a Calabi--Yau manifold $X$, an ample line bundle $L \to X$, and a basis $\{s_I\}_{I=1,\ldots,h^0(X,L^k)}$ of holomorphic sections of $L^k$, where $k$ is a positive integer, one considers the family of \K potentials
\begin{equation}
K_k(H) ~=~ \frac{1}{\pi k}\log\!\left(\sum_{I,J}H_{I J}\, s_I \overline{s_{J}}\right)\, ,
\label{eq:don_Ansatz}
\end{equation}
where $H$ is a positive-definite Hermitian matrix. In the context of Donaldson's algorithm, the matrix $H$ is determined, for fixed $k$, by imposing the \emph{balancing condition}, formulated as a fixed-point equation for a natural map (the $T$-map) on the space of Hermitian matrices. It has been proven in Ref.~\cite{Donaldson:2005hvr} that, as $k \to \infty$, the resulting sequence of balanced metrics converges to the unique Ricci-flat metric whose \K form lies in the class $2\pi\,c_1(L)$. The dependence of the metric on complex-structure moduli enters through the sections $s_I$ and through the coefficients $H_{I J}$ determined by the balancing procedure.
Despite these attractive features, practical applications of Donaldson’s algorithm face several limitations. First, achieving high numerical accuracy requires relatively large values of $k$, typically $k \gtrsim 6$, which significantly increases the computational cost~\cite{Douglas:2006rr} and implies a rather complicated form of Eq.~\eqref{eq:don_Ansatz}. More importantly for our purposes, the treatment of the \K moduli is structurally constrained. 

When $h^{1,1}(X)=1$, all \K classes are related by an overall rescaling, which affects only the total volume and not the shape of the Ricci-flat metric. In this case, approximating the metric in a fixed integral class suffices to recover the solution in any other class by a simple rescaling. For $h^{1,1}(X)>1$, however, selecting a specific \K class becomes more subtle. In principle one can vary the class by choosing a different line bundle $L$, but the resulting \K form $J$ is always tied to $c_{1}(L) \in H^{2}(X,\mathbb{Z})$.
On the other hand, the \K cone is continuous, and a general \K class can be parametrised as
\begin{equation}
J ~=~ \sum_{r=1}^{h^{1,1}(X)}t_r J_r ~\in~ H^{1,1}(X,\mathbb{R})~,
\label{eq:K_class_para}
\end{equation}
where $t_r$ are real \K parameters and the classes $J_r$ are a suitable basis of the second cohomology. Since $c_{1}(L)$ takes values in integral cohomology, Donaldson’s scheme samples only a discrete subset of \K classes. Varying the \K parameters in any way other than by an overall rescaling requires choosing a different line bundle and hence a different Ansatz. As a result, the approximate analytic metrics obtained at each level $k$ within Donaldson’s framework cannot incorporate explicit K\"ahler-moduli dependence, since the \K class is fixed by the choice of line bundle and therefore hard-coded in the Ansatz.

One might nevertheless attempt to retain the Ansatz~\eqref{eq:don_Ansatz} at fixed $k$ and determine the coefficients $H_{IJ}$ by directly optimising an approximation to the Ricci-flat condition, promoting them to functions of the \K parameters. However, the \K class of the metric obtained from~\eqref{eq:don_Ansatz} is determined entirely by the underlying polarisation and is independent of both $k$ and the specific choice of $H_{IJ}$. As a result, varying $H_{IJ}$ cannot move the metric within the \K cone, and the Ansatz is unable to realise a general parametrisation of the form~\eqref{eq:K_class_para}. We require a different Ansatz which allows for continuous K\"ahler class variations.

\subsection{Analytic Ansatz for the K\"ahler potential}

We now discuss an Ansatz for the \K potential capable of realising the parametrisation in Eq.~\eqref{eq:K_class_para}. For concreteness, we consider Calabi--Yau manifolds realised as (complete intersections of) hypersurfaces in ambient spaces $\mathcal{A}$, such as products of projective spaces or toric varieties. To simplify the discussion, we further assume that the K\"ahler cone of the Calabi--Yau manifold $X \subset \mathcal A$ descends from that of the ambient space. In other words, the classes generating $H^{1,1}(X)$ are obtained as pullbacks of divisor classes on $\mathcal A$. 

In this work we take $\mathcal A$ to be a product of complex projective spaces, $\mathcal A=\prod_{r=1}^{m}\mathbb P^{n_r}$, and denote by $i:X\hookrightarrow\mathcal A$ the inclusion of the Calabi--Yau hypersurface (or complete intersection). Throughout, we use $x$ to denote a generic point of $X$, represented whenever convenient by ambient-space coordinates restricted to the Calabi--Yau manifold. Let $H_r\in H^{1,1}(\mathcal A,\mathbb Z)$ denote the hyperplane class of the $r$-th projective factor. Then the classes $J_r \;:=\; i^{*} H_r \;\in\; H^{1,1}(X,\mathbb Z)$
span $H^{1,1}(X,\mathbb R)$ and hence the \K cone of $X$. A general \K class on $X$ can then be written as
\begin{equation} \label{Jexp}
J ~=~ \sum_{r=1}^{m} t_r \,J_r ~\in~ H^{1,1}(X,\mathbb R)\, ,
\end{equation}
with \K parameters $t_r$ which we denote collectively as $t=(t_1,\ldots,t_m)$. In many cases of interest (including the examples presented below) the K\"ahler cone is characterised by $t_r>0$ for all $r$. We also introduce the Calabi--Yau volume
\begin{equation}
V(t)~=~ \frac{1}{3!}\int_X J^3 ~=~ \frac{1}{6}\sum_{r,s,u=1}^{m} d_{rsu}\, t_r t_s t_u \;,\qquad
d_{rsu} ~=~ \int_X J_r \wedge J_s \wedge J_u
\end{equation}
where $d_{rsu}$ are the triple intersection numbers of $X$.

Once a K\"ahler class is fixed, any two K\"ahler forms in the same cohomology class differ by a $\partial\bar\partial$–exact term. The \K potential $K$ associated with the Ricci-flat metric can therefore be written as
\begin{equation}\label{Kphi}
K ~=~ K_{\mathrm{FS}} + \phi \, ,\qquad 
K_{\mathrm{FS}} ~=~ \sum_{r=1}^{m} \frac{t_r}{\pi}\, \log\!\left(\sum_{a=0}^{n_r} |x^{(r)}_a|^2 \right) \, ,
\end{equation}
where $\phi$ is a globally defined real-valued function on the Calabi--Yau manifold, which is unique up to an additive constant and $K_{\rm FS}$ is the Fubini-Study K\"ahler potential. The ambiguity in $\phi$ can be removed by imposing the condition 
\begin{equation}
\int_X \phi \, {\rm dVol}_{\rm FS} ~=~ 0 \, ,
\label{eq:norm_cond}
\end{equation}
where ${\rm dVol}_{\rm FS}$ is the measure associated to the Fubini-Study K\"ahler potential $K_{\rm FS}$.

In the neural-network approach of Ref.~\cite{Larfors:2022nep}, implemented in the {\sf cymetric} package, the function $\phi$ is represented by a neural network. This network is trained by minimising a loss function encoding the Ricci-flatness condition in the form of the complex Monge--Amp\`ere equation as well as conditions to ensure that the K\"ahler class of the Ricci-flat metric is indeed given by Eq.~\eqref{Jexp}. (The latter is, of course, mathematically guaranteed by $\phi$ being a function but this has to be enforced in a numerical approach.) In the examples studied in this paper and discussed below, we use the {\sf cymetric} package to compute the function $\phi$ numerically. 

We also consider an explicit analytic Ansatz for $\phi$ based on a finite-dimensional space of sections of an ample line bundle 
$L=\mathcal O_{X}(k)$, where $k=(k_1,\dots,k_m)$,
and denote by $(s_I)_{I=1,\dots,h^0(X, L)}$ a basis of global holomorphic sections of $L$. For fixed multi-degree $k$, our Ansatz reads
\begin{equation}
\phi(x,t) ~=~ V(t)^{1/3}\,\frac{\sum_{I,J} \alpha_{IJ}(t_1/t_m,\ldots,t_{m-1}/t_m)\,s_I(x)\,\overline{s_J(x)}} {\prod_{r=1}^{m} \left(\sum_{a=0}^{n_r} |x^{(r)}_a|^2\right)^{k_r}} \, ,
\label{eq:phi_general}
\end{equation}
where the coefficients $\alpha_{IJ}$ form a Hermitian matrix and depend only on dimensionless \K moduli ratios, something that can be understood as follows. 
Under a rescaling of the K\"ahler parameters $t_r\to\lambda t_r$, the K\"ahler class and the corresponding Ricci-flat metric scale as $J\to\lambda J$ and $g\to\lambda g$. Since the metric is given by $g_{i\bar j}=\partial_i\partial_{\bar j}K$, the K\"ahler potential must therefore scale linearly,
$K(x,\lambda t)=\lambda K(x,t)$ (up to a K\"ahler transformation). Because $K_{\rm FS}$ already scales linearly in $t$, the correction
$\phi$ must satisfy $\phi(x,\lambda t)=\lambda\,\phi(x,t)$. This homogeneity is implemented in Eq.~\eqref{eq:phi_general} through the factor $V(t)^{1/3}$. As such, the remaining dependence of the coefficients $\alpha_{IJ}$ must only depend on dimensionless ratios, which we take to be $t_r/t_m$, for $r=1,\ldots , m-1$. 

By construction, the terms appearing in Eq.~\eqref{eq:phi_general} define globally well-defined functions on $X$ (and in fact extend to the ambient space $\mathcal A$), independent of the choice of local trivialisation of $L$. This Ansatz may be interpreted as a spectral truncation to low-frequency modes of the scalar Laplacian, with a cut-off determined by the multi-degree $k$ (see Ref.~\cite{Anderson:2023viv} for a detailed discussion).

Sections of $L$ on $X$ are obtained by restricting ambient sections. More precisely, if $X$ is defined as the zero locus of a single polynomial $P$ of multi-degree $(n_1+1,\dots,n_m+1)$, we have
\begin{equation}
\Gamma(X,L) \;\cong\; \Gamma(\mathcal A,\mathcal O_{\mathcal A}(k)) \big/ \bigl(P\cdot \Gamma(\mathcal A,\mathcal O_{\mathcal A}(k-d))\bigr),
\end{equation}
Consequently, the matrix $(\alpha_{IJ})$ in Eq.~\eqref{eq:phi_general} contains $(h^0(X,L))^2$ real parameters (subject to the Hermiticity condition).
For sufficiently small multi-degree $k$, that is when at least one component of the line-bundle multi-degree satisfies $k_r\leq n_r$, the restriction map from ambient sections to sections on $X$ is injective, so that
\begin{equation}
\Gamma(X,L)\cong \Gamma(\mathcal A,\mathcal O_{\mathcal A}(k)).
\end{equation}
and the ambient monomials provide a basis of sections on $X$. 
This is, in fact, the case for  the examples considered below, since we work with relatively small values of the line-bundle degrees. For larger values of $k$, however, sections differing by multiples of the defining polynomial are identified upon restriction to $X$.

\subsection{Data-driven approach}\label{sec:data}
Our approach to computing approximate Ricci-flat metrics based on the above Ansatz is as follows.
\begin{itemize}
\item For a given point $t$ in K\"ahler moduli space with $V(t)=1$, we use the {\sf cymetric} package~\cite{Larfors:2021pbb} to compute the Ricci-flat metric and its associated K\"ahler potential numerically.
\item The function $\phi$ and, hence, the coefficients $\alpha_{IJ}$ in Eq.~\eqref{eq:phi_general} are determined by a fit to the numerical result for the K\"ahler potential.  We impose Eq.~\eqref{eq:norm_cond} to ensure that $\phi$ is unique.
\item The above two steps are repeated for many values $t$ in K\"ahler moduli space with $V(t)=1$. In this way, we gain information about the K\"ahler moduli dependence of the coefficients $\alpha_{IJ}$.
\item Finally, we apply symbolic regression, specifically {\sf PySR}~\cite{PySR}, to the above data to find analytic expressions for the coefficients $\alpha_{IJ}$ as functions of the ratios $t_r/t_m$, $1\leq r\leq m-1$. Inserting these expressions into Eqs.~\eqref{Kphi} and~\eqref{eq:phi_general} then provides the analytic K\"ahler potential for an approximate Ricci-flat metric. 
\end{itemize}
In principle, one could bypass the neural-network stage altogether by applying symbolic regression directly to sampled points on the Calabi--Yau manifold using the Ansatz above and a loss function that encodes the Ricci-flatness condition. In this approach, the analytic dependence of the coefficients $\alpha_{IJ}$ on the moduli ratios $t_r/t_m$ would be obtained by directly minimising the Ricci-flatness loss, eliminating the need to first learn the K\"ahler potential numerically, fit the Ansatz, and subsequently apply symbolic regression. However, similar approaches have already been explored and appear to be computationally expensive~\cite{Mirjanic:2024gek}. For this reason, we shall stick with the procedure illustrated in the bullet-list above and apply it to two Calabi--Yau manifolds that admit nontrivial discrete symmetries, denoted by $G$. Such a $G$-action on $X$ induces a linear action on the K\"ahler moduli space, that is, a representation
\begin{equation}
\rho:G \longrightarrow {\rm GL}\bigl(H^{1,1}(X,\mathbb R)\bigr),
\end{equation}
which preserves the K\"ahler cone. Its kernel $N:=\ker(\rho)$ is a normal subgroup of $G$, consisting of those symmetries that act trivially on the K\"ahler structure. Hence $\rho(G)\cong G/N$, and we obtain the short exact sequence
\begin{equation}
1 \longrightarrow N \longrightarrow G \longrightarrow G/N \longrightarrow 1.
\end{equation}
If this extension splits, then $G$ can be written as a semidirect product
\begin{equation}\label{GH}
G = N \rtimes H\, , \qquad H \cong G/N\, .
\end{equation}
For intuition, note that if $G$ acts freely and preserves the holomorphic three-form, then the quotient $X/G$ is again a Calabi--Yau manifold. The pullback of its Ricci-flat metric is therefore $G$-invariant. More generally, the same conclusion follows directly from the uniqueness part of Yau's theorem~\cite{Yau1978}, without any assumption that the action be free.\footnote{
Let $g:X\to X$ be a holomorphic automorphism, and let $\omega_t$ denote the unique Ricci-flat K\"ahler form in the K\"ahler class labelled by $t=(t_1,\dots,t_m)$. By definition of the induced action on K\"ahler classes,
\[
[g^*\omega_t]=[\omega_{\rho(g)t}]\, .
\]
Moreover, $g^*\omega_t$ is again Ricci-flat. Hence $g^*\omega_t$ is the Ricci-flat K\"ahler form in the class $[\omega_{\rho(g)t}]$, and by the uniqueness part of Yau's theorem,
\[
g^*\omega_t=\omega_{\rho(g)t}\, .
\]
In particular, if $g\in N=\ker\rho$, then $g^*\omega_t=\omega_t$. This argument does not require the action of $g$ to be free.} At this stage, one could reduce the number of independent coefficients in $(\alpha_{IJ})$ by imposing the symmetry on $\phi$ and organising the monomials into $N$-invariant combinations. Instead, we begin with a symmetry-agnostic Ansatz and use it as a consistency check of the procedure, verifying numerically that the fitted coefficients respect the expected symmetry relations. Having established this agreement, we then impose $G$-invariance on the Ansatz for $\phi$ and repeat the fit using the resulting symmetry-constrained Ansatz.

To build this constrained Ansatz, we then identify the $N$-singlets $I_i(x)$ in $\Gamma(X,L)\otimes \Gamma(X,L)^*$, that is, specific linear combinations of $s_I(x)\,\overline{s_J(x)}$, and rewrite the Ansatz~\eqref{eq:phi_general} as
\begin{equation}\label{ansGinv}
\phi(x,t) ~=~ V(t)^{1/3}\, \frac{\displaystyle\sum_{i} \alpha_i(t_1/t_m,\ldots,t_{m-1}/t_m)\,I_i(x)} {\displaystyle\prod_{r=1}^{m}\left(\sum_{a=0}^{n_r}|x_a^{(r)}|^2\right)^{k_r}} \, ,
\end{equation}
where the sum in the numerator runs over all $N$-singlets $I_i$.
Since $N$ acts trivially on the K\"ahler moduli, the coefficients $\alpha_i$ are automatically $N$-invariant functions of $t$. On the other hand, the quotient group $H$ acts non-trivially both on the moduli and on the singlets,
\begin{equation}
I_i(hx)=r(h)_{ij}\,I_j(x)\, ,
\end{equation}
for some representation
\begin{equation}
r:H\to \mathrm{End}\bigl((\Gamma(X,L)\otimes\Gamma(X,L)^*)^N\bigr)\, .
\end{equation}
Here $(\Gamma(X,L)\otimes\Gamma(X,L)^*)^N$ denotes the subspace of $N$-invariant elements in $\Gamma(X,L)\otimes\Gamma(X,L)^*$.
In order for the Ansatz~\eqref{ansGinv} to respect the symmetry, the coefficients must compensate this transformation,
\begin{equation}\label{alphatrans}
\alpha_i(\rho(h)t)=(r(h)^{-1})_{ji}\,\alpha_j(t)\, .
\end{equation}
With this $G$-symmetric set-up, we repeat the four steps in the above bullet-point list using the Ansatz~\eqref{ansGinv} instead of Eq.~\eqref{eq:phi_general}, thereby determining explicit expressions for the coefficients $\alpha_i$ as functions of the K\"ahler moduli ratios. The non-trivial transformation laws~\eqref{alphatrans} then provide a useful and non-trivial consistency check on the numerical results, as illustrated in the examples below.

\subsection{Examples with two K\"ahler moduli}
For simplicity we will consider examples of Calabi--Yau manifolds that are defined as hyper-surfaces in products  of two projective spaces. In this sub-section we will specialise our general approach to this subset of examples.

More specifically, we are working with ambient spaces of the form $\mathcal{A}=\mathbb{P}^{n_1} \times \mathbb{P}^{n_2}$ and homogeneous coordinates $x=(x_0,\ldots ,x_{n_1})$ and $y=(y_0,\ldots ,y_{n_2})$ for the two projective factors. Within such ambient spaces we consider Calabi--Yau hyper-surfaces defined as the zero locus of a homogeneous polynomial $P=P(x,y)$ of bi-degree $(n_1+1,n_2+1)$. The K\"ahler class $J$ is parametrized by
\begin{equation}
 J=t_1 J_1+t_2 J_2\; ,
\end{equation}
with K\"ahler parameters $t=(t_1,t_2)$
and Eq.~\eqref{Kphi} specializes to 
\begin{equation}
K(x,y,t_1,t_2) ~=~ \frac{t_1}{\pi}\, \log \left(\sum_{a=0}^{n_1} |x_a|^2 \right)     ~+~ \frac{t_2}{\pi}\, \log \left(\sum_{a=0}^{n_2} |y_a|^2 \right) +\phi(x,y,t_1,t_2)~. 
\label{eq:FS_metric}
\end{equation}
For a line bundle $L={\cal O}_X(k)$, where $k=(k_1,k_2)$, the Ansatz for $\phi$ in Eq.~\eqref{eq:phi_general} becomes
\begin{equation}
\phi(x,y,t_1,t_2) ~=~ V(t_1,t_2)^{1/3}\,
\frac{\sum_{I,J}
\alpha_{IJ}(t_{12})\, s_I(x,y)\,\overline{s_J(x,y)}}{\bigl(\sum_a |x_a|^2\bigr)^{k_1}\bigl(\sum_a |y_a|^2\bigr)^{k_2}} \, .
\label{eq:phi}
\end{equation}
with the coefficients $\alpha_{IJ}=\alpha_{IJ}(t_{12})$  depending only on the K\"ahler modulus ratio $t_{12}=t_1/t_2$. Similarly, the $G$-invariant Ansatz~\eqref{ansGinv} specialises to 
\begin{equation}
\phi(x,y,t_1,t_2) ~=~ V(t_1,t_2)^{1/3}\,
\frac{\sum_{i}
\alpha_{i}(t_{12})\, I_i(x,y)}{\bigl(\sum_a |x_a|^2\bigr)^{k_1}\bigl(\sum_a |y_a|^2\bigr)^{k_2}} \, .
\label{eq:phiGinv}
\end{equation}

\section{Ricci-flat metrics on the bi-cubic}
\label{sec:bicubic}
In this section, we tackle our first example, a bi-cubic hyper-surface in the ambient space $\mathcal{A}=\mathbb{P}^2\times\mathbb{P}^2$ with homogeneous coordinates $x=(x_0,x_1,x_2)$ and $y=(y_0,y_1,y_2)$ for the two projective factors. Bi-cubic Calabi--Yau manifolds $X$ are defined as the zero locus in $\mathcal{A}$ of a homogeneous bi-degree $(3,3)$ polynomial $P=P(x,y)$ and they have Hodge numbers $h^{1,1}(X)=2$ and $h^{2,1}(X)=83$. We will focus on a single point in complex structure moduli space, with a large symmetry group $G$ to be discussed shortly, and explore the dependence of the Ricci-flat K\"ahler potential on the two K\"ahler parameters $t=(t_1,t_2)$. The volume takes the form 
\begin{equation}\label{eq:bcvol}
 V(t_1,t_2)=\frac{3}{2}t_1t_2(t_1+t_2)\; ,
\end{equation} 
and the K\"ahler cone is characterised by $t_1>0$ and $t_2>0$.

\subsection{The choice of bicubic threefold}\label{sec:bc}
To simplify our calculations we would like to consider a bi-cubic with a large symmetry group. To this end we introduce $\omega=e^{2\pi i/3}$ and a number of groups generated as follows.
\begin{equation}
    \begin{array}{rclcl}
        (\mathbb{Z}_3^{(x)})_a \!\!\!\!&:& x_a\mapsto \omega x_a&,& x_b\mapsto x_b\;\;\mbox{for}\;\; b\neq a\,,\;\; y_c\mapsto y_c\;\;\mbox{for}\;\; c=0,1,2\\[2mm]
        (\mathbb{Z}_3^{(y)})_a \!\!\!\!&:& y_a\mapsto \omega y_a&,& y_b\mapsto y_b\;\;\mbox{for}\;\; b\neq a\,,\;\; x_c\mapsto x_c\;\;\mbox{for}\;\; c=0,1,2\\[2mm]
        S_3^{(x,y)}&:& x_a\mapsto x_{\sigma(a)}&,& y_a\mapsto y_{\sigma(a)}\;\;\mbox{where}\;\;\sigma\in S_3\\[2mm]
        \mathbb{Z}_2^{(x,y)}&:& x_a\mapsto y_a&,& y_a\mapsto x_a\;\;\mbox{for}\;\; a = 0,1,2
    \end{array}
\end{equation}
The transformations above act on homogeneous coordinates and therefore define projective symmetries of $\mathbb P^2\times\mathbb P^2$. In particular, the diagonal phase rotation $(x_0,x_1,x_2)\mapsto (\omega x_0,\omega x_1,\omega x_2)$ is trivial on the first $\mathbb P^2$, and similarly for the second. Hence the effective phase symmetry is $(\mathbb Z_3^{(x)})^2 \times (\mathbb Z_3^{(y)})^2$. Together with the simultaneous permutation symmetry $S_3^{(x,y)}$ and the exchange symmetry $\mathbb Z_2^{(x,y)}$, this yields the effective projective symmetry group
\begin{equation}
G \cong \Bigl((\mathbb Z_3^{(x)})^2 \times (\mathbb Z_3^{(y)})^2\Bigr)\rtimes (S_3\times \mathbb Z_2)~,
\end{equation}
of order $|G|=4^2\cdot 6\cdot 2 = 972$. Only the $\mathbb Z_2^{(x,y)}$ factor acts non-trivially on the K\"ahler class, exchanging the two K\"ahler moduli $t_1\leftrightarrow t_2$. Accordingly, we write 
\begin{equation}
G=N\rtimes H, \qquad
N=\Bigl((\mathbb Z_3)^2_x\times (\mathbb Z_3)^2_y\Bigr)\rtimes S_3, \qquad H\cong \mathbb Z_2.
\end{equation}

The sub-space of $G$-invariants in the space of homogeneous polynomials of bi-degree $(3,3)$ is two-dimensional and is spanned by
\[
P_1=\sum_a x_a^3y_a^3,\qquad P_2=\sum_{a\neq b}x_a^3y_b^3\;.
\]
Hence, the most general defining polynomial for a $G$-invariant bi-cubic can be written as $P=\lambda_1P_1+\lambda_2P_2$ and this represents a one-parameter family in complex structure moduli space since an overall scaling of $P$ defines the same manifold. For definiteness, we choose a particular point within this one-parameter family, namely
\begin{equation}
P=10\,P_1+ 3\, P_2\; .
\label{eq:P_def_bicubic}
\end{equation}
It can be checked that this polynomial defines a smooth Calabi--Yau space.

\subsection{Invariants}
Having identified the effective projective symmetry group $G$, we now exploit this symmetry to simplify the analytic Ansatz for the K\"ahler potential. In principle the coefficients $\alpha_{IJ}$ in Eq.~\eqref{eq:phi} parametrise all functions constructed from products $s_I\overline{s_J}$ of sections of $L$. 
For our fit of the numerical results to the Ansatz~\eqref{eq:phi} we will focus on the line bundle $L={\cal O}_X(2,2)$ on the bi-cubic. Its space of sections has dimension $h^0(X,{\cal O}_X(2,2))=36$ so that the matrix $(\alpha_{IJ})$ in Eq,~\eqref{eq:phi} has size $36\times 36$. 

This is rather large for writing down explicit analytic expression. However, if our Ansatz respects the symmetry $G$ of our chosen bi-cubic -- and this will be verified numerically -- then we can work with the Ansatz~\eqref{eq:phiGinv}, focusing on the $N$-invariant sub-space of  $\Gamma(X,{\cal O}_X(2,2))\times \Gamma(X,{\cal O}_X(2,2))^*$. It turns out this sub-space is eight-dimensional and it is spanned by the invariants
\begin{equation}
\begin{array}{rclcrcl}
    I_0 &=& \ds \frac{1}{3}\sum_{a < b} \, |x_a|^2 \, |x_b|^2 \,|y_a|^2 \, |y_b|^2&&
    I_1 &=&\ds \frac{1}{3}\sum_{\substack{a \neq b,c \\ b<c}} \, |x_a|^4 \, |y_b|^2 \, |y_c|^2\\[8mm]
    I_2 &=& \ds\frac{1}{6}\sum_{a \neq b} \, |x_a|^4 \, |y_a|^2 \, |y_b|^2&&
    I_3 &=& \ds\frac{1}{6}\sum_{a \neq b} \, |x_a|^4 \, |y_b|^4\\[8mm] 
    I_4 &=&\ds\frac{1}{3}\sum_{a} \, |x_a|^4 \, |y_a|^4&&
    I_5 &=&\ds \frac{1}{6}\sum_{\substack{a, b, c \\ {\rm distinct}} } \, |x_a|^2 \, |y_a|^2 \, |x_b|^2 \, |y_c|^2\\[8mm]
    I_6 &=&\ds \frac{1}{6}\sum_{a \neq b} \, |y_a|^4 \, |x_a|^2 \, |x_b|^2&&
    I_7 &=&\ds \frac{1}{3}\sum_{\substack{a \neq b,c \\ b<c}} \, |y_a|^4 \, |x_b|^2 \, |x_c|^2~.
    \end{array}
    \label{eq:singlets}
\end{equation}
Under the generator of the group $H\cong\mathbb{Z}_2^{(x,y)}$ these singlets transform as $I_1\leftrightarrow I_7$ and $I_2\leftrightarrow I_6$ with $I_0,I_3,I_4,I_5$ invariant. From Eq.~\eqref{alphatrans} we, therefore, expect the coefficients $\alpha_i$ to transform as
\begin{equation}\label{alpharel}
 \alpha_1\left(\frac{t_1}{t_2}\right)=\alpha_7\left(\frac{t_2}{t_1}\right)\;,\quad
 \alpha_2\left(\frac{t_1}{t_2}\right)=\alpha_6\left(\frac{t_2}{t_1}\right)\;,\quad
 \alpha_i\left(\frac{t_1}{t_2}\right)=\alpha_i\left(\frac{t_2}{t_1}\right)\;\mbox{ for }\;i=0,3,4,5\; .
\end{equation}
As mentioned before, verifying these transformation laws from our numerical results provides a useful and non-trivial check.

\subsection{Numerical results}
Following the approach outlined in Section~\ref{sec:data}, we first need to compute the Ricci-flat K\"ahler potential numerically, for a number of K\"ahler parameter values $t=(t_1,t_2)$. Since the dependence on the overall Calabi--Yau volume $V(t)$ (for the bi-cubic given in Eq.~\eqref{eq:bcvol}) is already explicit in the Ansatz~\eqref{eq:phi}, it is sufficient to do this for $t$-values with $V(t)=1$. Specifically, we select the set of K\"ahler moduli values
\begin{equation}\label{t12val}
    \left\{t=(t_1,t_2)\,~\bigg|~\,V(t)=1\,,\; \frac{t_1}{t_2}=\frac{p}{q}~\mbox{ where } p,q=1,\ldots  14\mbox{ and gcd}(p,q)=1\right\}
\end{equation}
resulting in a total of 127 distinct values of $(t_1,t_2)$.
\begin{figure}[h!]
    \centering    \includegraphics[width=0.395\textwidth]{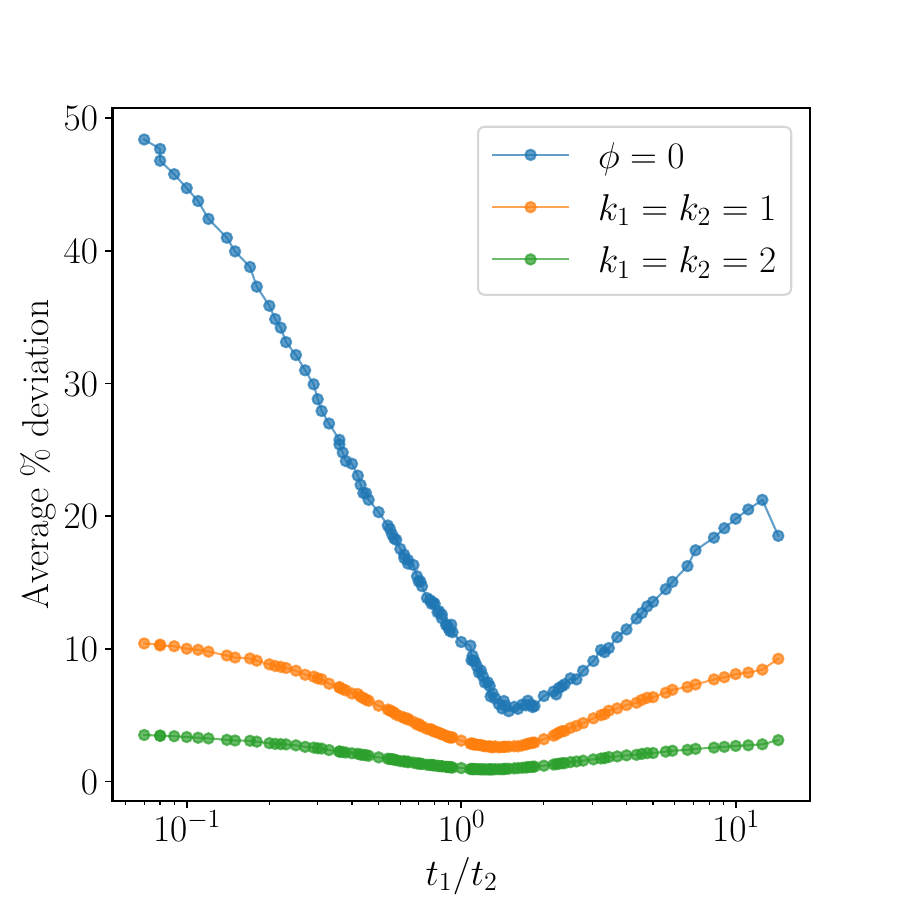}
    \caption{Average percentage deviation of the K\"ahler potential, $\langle |K_{\rm NN}-K|/K_{\rm NN}\rangle$, as a function of $t_1/t_2$, for different ans\"atze. The blue points correspond to the pure Fubini--Study potential ($\phi=0$), while the orange and green points show fits obtained from the Ansatz in Eq.~\eqref{eq:phi} with $(k_1,k_2)=(1,1)$ and $(k_1,k_2)=(2,2)$, respectively. Increasing the truncation order leads to a significant improvement in agreement with the numerically learned \K potential. Averaged over $t_1/t_2$, the deviations are $17.1\%$, $5.3\%$, and $1.8\%$ for the three cases, respectively.}
    \label{fig:bicubic_avg_deviation}
\end{figure}

For each choice of the \K parameters in the above set, we use the point generator provided by the \texttt{cymetric} package~\cite{Larfors:2021pbb} to sample the manifold defined by the polynomial~\eqref{eq:P_def_bicubic} with $N_{\rm pts}=100{,}000$ points and train a fully-connected neural network (width 64, depth 3, GeLU activation) to learn the function~$\phi$. The network is randomly initialised only at the first value of $t_{12}=t_1/t_2$ considered. For subsequent calculations we proceed to the closest $t_{12}$ value and initialise the neural network with the optimal parameter obtained in the previous run. This warm-start strategy leads to a substantial reduction in training time. In particular, following an initial training run of 240 epochs, all subsequent runs at different values of $t_{12}$ require only 10 epochs to reach a loss comparable to, or smaller than, that achieved for the first $t_{12}$ value.   
After imposing the normalisation condition in Eq.~\eqref{eq:norm_cond}, we determine the best-fit values of the coefficients $\alpha_{IJ}$ using the Ansatz in \eqref{eq:phi} for $(k_1,k_2) = (1,1)$ and $(k_1,k_2) = (2,2)$.  Fig.~\ref{fig:bicubic_avg_deviation} displays the percentage deviation of the analytic expression obtained from the fit relative to the numerical \K potential learned by the neural network. We denote the analytic \K potential by~$K$ and its numerical counterpart by $K_{\rm NN}$. Averaging over the range in $t_1/t_2$, the pure Fubini-Study K\"ahler potential (blue curve in the right panel of Fig.~\ref{fig:bicubic_avg_deviation}) provides a relatively poor approximation with a deviation of $17$\%. Already for $(k_1,k_2)=(1,1)$ this improves to about $5$\% and for $(k_1,k_2)=(2,2)$ to about $2$\%.

\begin{figure}[h!]
    \centering
    \includegraphics[width=0.328\textwidth]{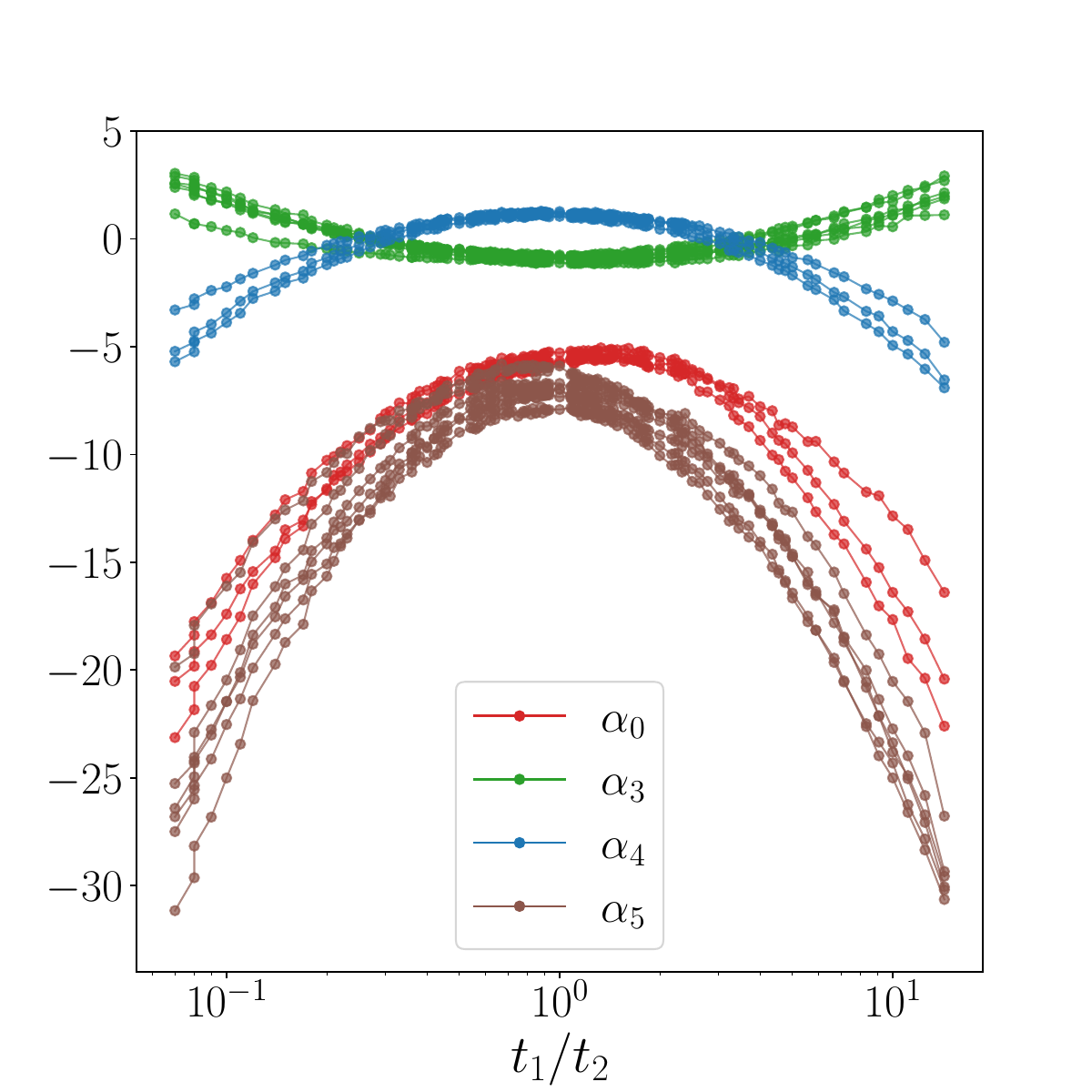}
    \includegraphics[width=0.328\textwidth]{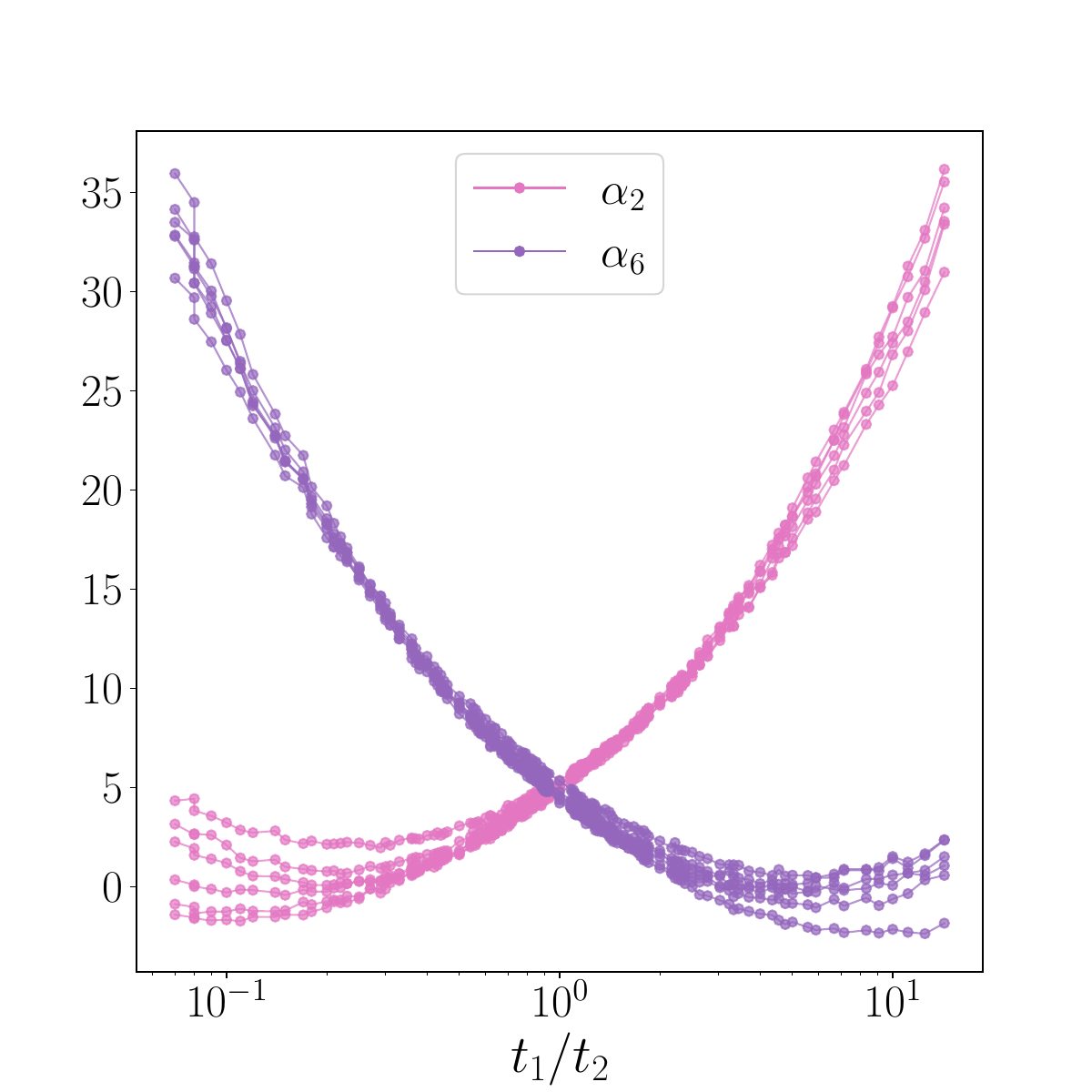}
    \includegraphics[width=0.328\textwidth]{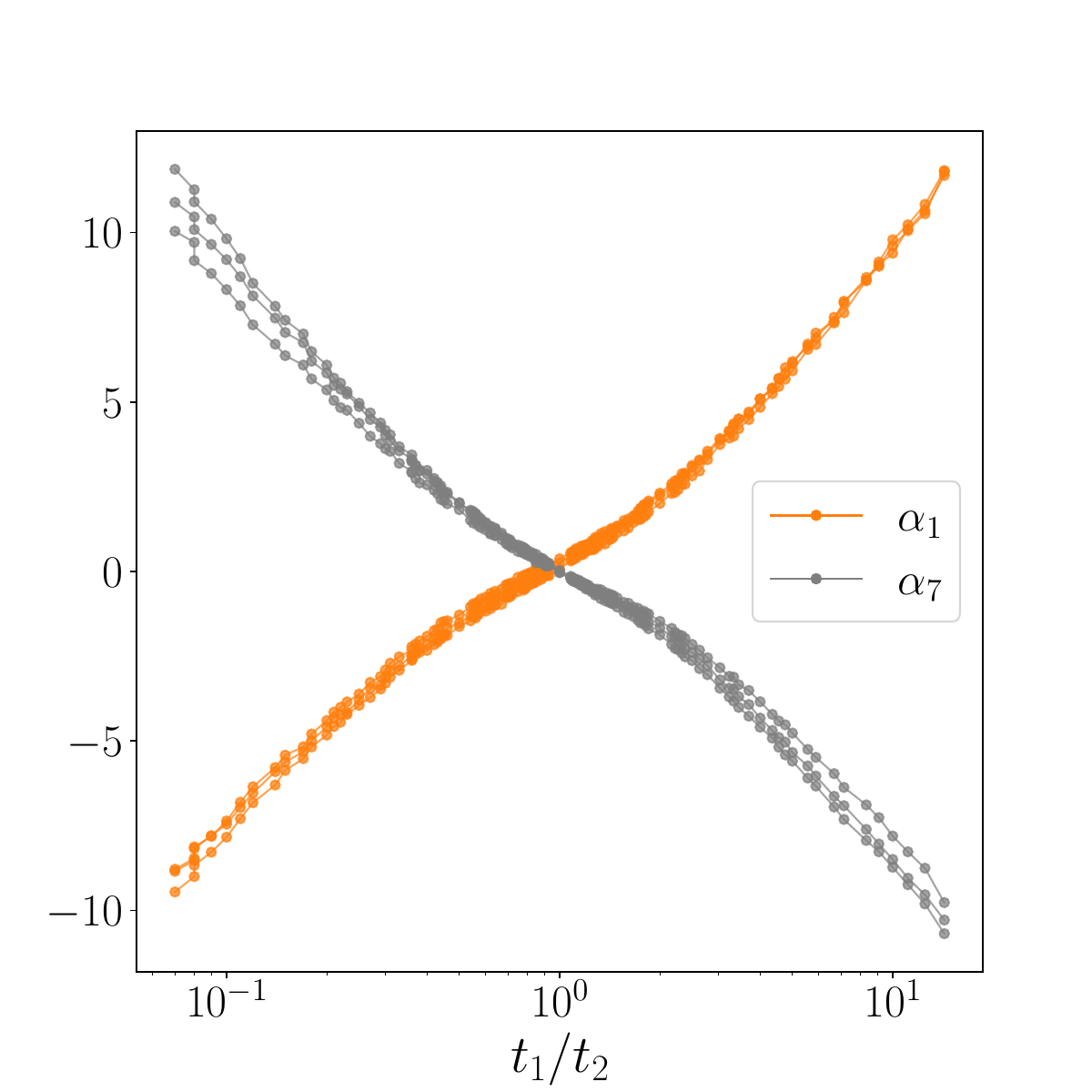}

    \caption{Best-fit coefficients as functions of $t_1/t_2$ for the $(k_1,k_2)=(2,2)$ truncation. Out of the $36\times 36$ entries of the Hermitian matrix $\alpha_{IJ}$, only $36$, the diagonal entries, are non-vanishing within numerical accuracy; as expected, these naturally organise into eight groups,  which are displayed in different colours. Coefficients corresponding to monomials that belong to the same $H$-singlet are shown with the same colour and labelled as $\alpha_0,\ldots,\alpha_7$. The left panel displays coefficients associated with singlets invariant under $x \leftrightarrow y$, while the remaining panels show pairs of coefficients associated with singlets that are exchanged under $x \leftrightarrow y$, exhibiting the expected crossing behaviour at $t_1=t_2$. 
}
    \label{fig:bicubic_alpha_plots}
\end{figure} 

The functions in~\eqref{eq:phi} form a complete basis on the ambient projective space in the limit $(k_1,k_2)\to\infty$, so increasing the truncation order should improve the accuracy further. However, to achieve this in practice the numerical calculation has to be carried out at the same level of accuracy. In the present paper we will not attempt to carry this out beyond the degree $(k_1,k_2)=(2,2)$.

\begin{figure}[h!]
    \centering
    \includegraphics[width=0.328\textwidth]{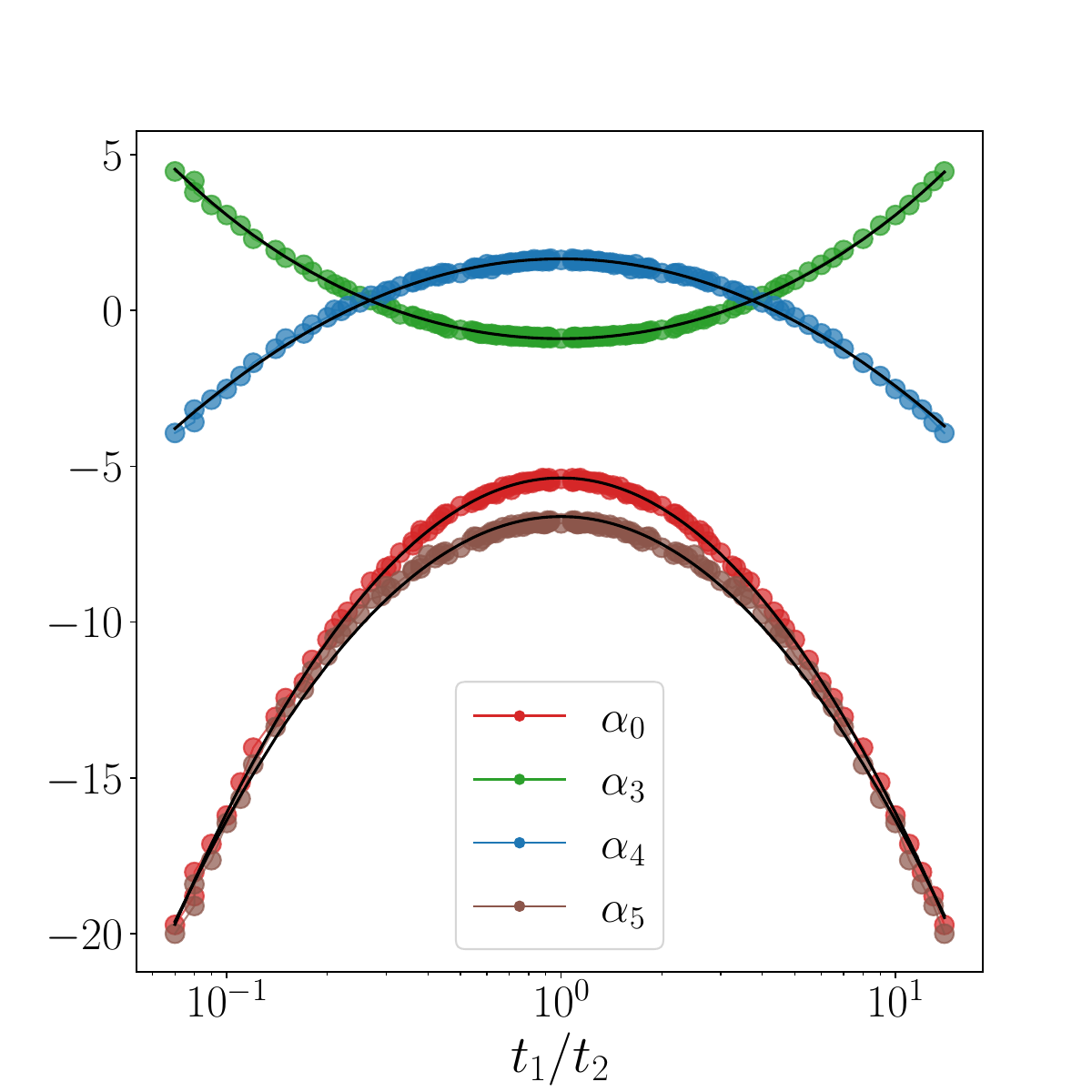}
    \includegraphics[width=0.328\textwidth]{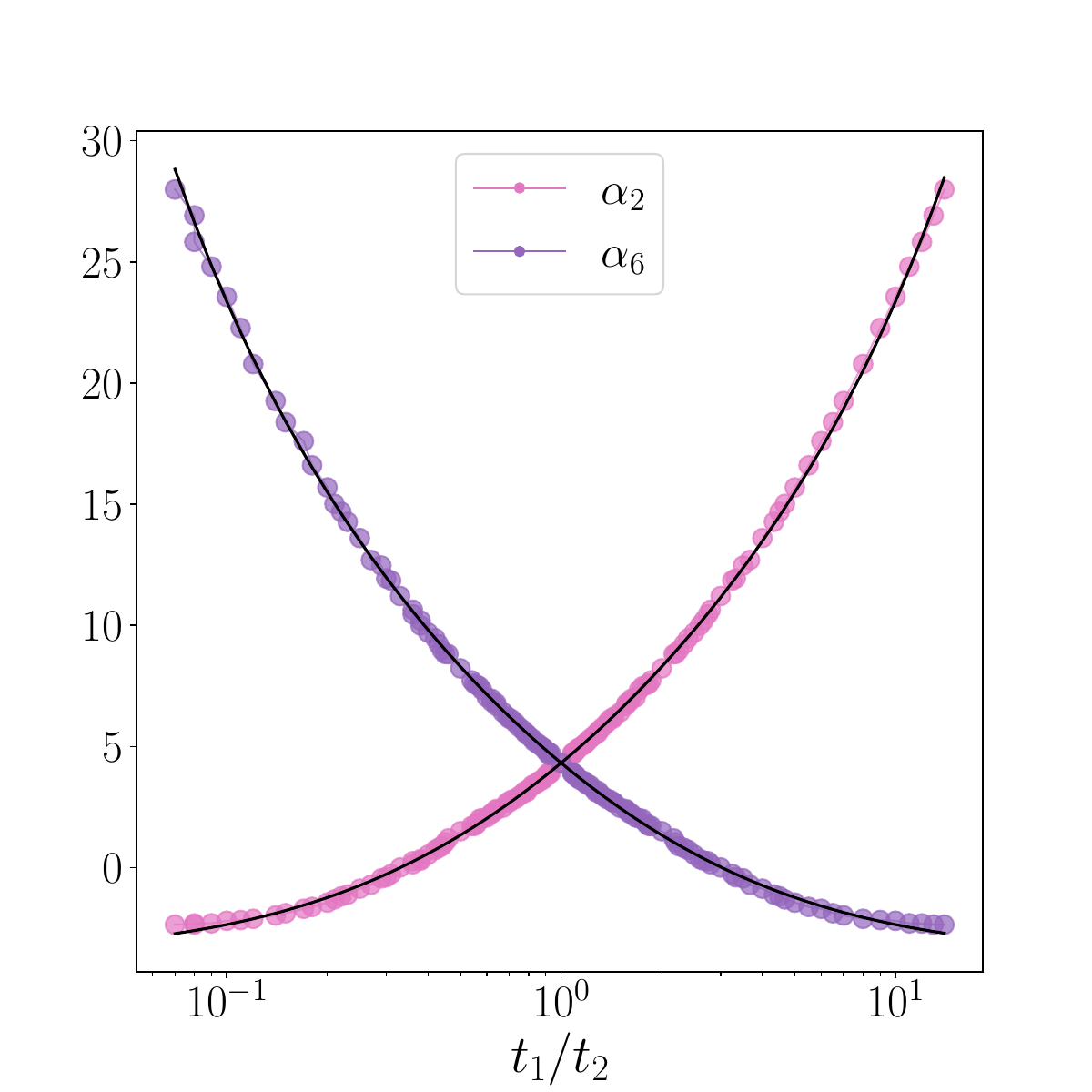}
    \includegraphics[width=0.328\textwidth]{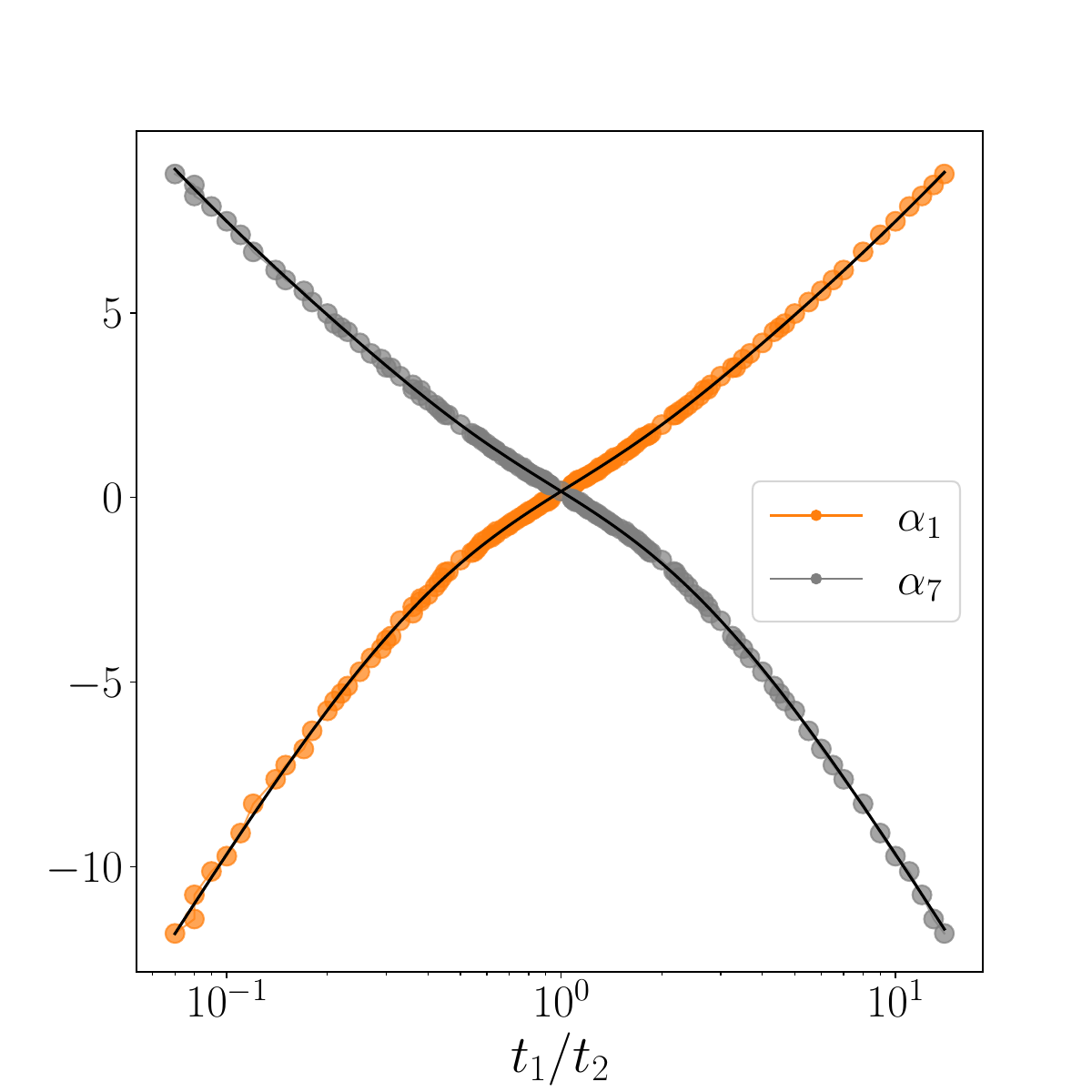}

    \caption{Best-fit numerical values of the coefficients $\alpha_i$ in the Ansatz~\eqref{eq:phiGinv} as functions of the moduli ratio $t_1/t_2$, obtained for $(k_1,k_2)=(2,2)$.}
\label{fig:bicubic_alpha_plots_symmetrised}
\end{figure}

Fig.~\ref{fig:bicubic_alpha_plots} shows the best-fit numerical values of the coefficients $\alpha_{IJ}$ for the case $(k_1,k_2)=(2,2)$. As discussed above, for this choice, the coefficients $\alpha_{IJ}$ in the Ansatz~\eqref{eq:phi} form a hermitian matrix  of size $36\times 36$, since $h^0(X,\mathcal O_X(2,2))=36$. The fit indicates that the off-diagonal entries are strongly suppressed relative to the diagonal ones. Curves in Fig.~\ref{fig:bicubic_alpha_plots} displayed with the same colour indicate coefficients that naturally group together and exhibit the same dependence on $t_1/t_2$ up to numerical uncertainties. These observations are in line with the theoretical expectations and provide a solid consistency check on the numerical fit and order truncation. We then proceed by writing our Ansatz in terms of the eight invariants $I_i$ defined in Eq.~\eqref{eq:singlets}. Fitting the data to this Ansatz leads to numerical results for the eight coefficients $\alpha_i=\alpha_i(t_1/t_2)$ which are shown in  Fig.~\ref{fig:bicubic_alpha_plots_symmetrised}. It is evident from these curves and can be checked quantitatively that these numerical results for $\alpha_i$ do indeed satisfy the expected symmetry relations in Eq.~\eqref{alpharel}.

\begin{table}[!ht]
\centering
\begin{tabular}{c l}
\hline
Coefficient & Fit expression \\
\hline
\vspace{0.01cm}\\
$\alpha_0$ &
$-5.37 - 2.03\, \log^2\!\left(t_{12}\right)$
\\[6pt]

$\alpha_1$ &
$3e^{-2.142\, t_{12}}\!\left(1.334 + \log\!\left(t_{12}\right)\right)
+ 3\log\!\left(0.869\, t_{12} + 0.034\, t_{12}^2\right)$
\\[6pt]

$\alpha_2$ &
$1.39 + 0.79\, t_{12} + 6\log\!\left(0.430 + t_{12}\right)$
\\[6pt]

$\alpha_3$ &
$-0.90 + 0.69\, \log^2\!\left(t_{12}\right)
+ 0.012\, \log^4\!\left(t_{12}\right)$
\\[6pt]

$\alpha_4$ &
$1.66 - 0.77\, \log^2\!\left(t_{12}\right)$
\\[6pt]

$\alpha_5$ &
$-6.61 - 1.84\, \log^2\!\left(t_{12}\right)$
\\[6pt]

$\alpha_6$ &
$1.39 + 0.8\, t_{21} + 6\log\!\left(0.430 + t_{21}\right)$
\\[6pt]

$\alpha_7$ &
$3e^{-2.142\, t_{21}}\!\left(1.334 + \log\!\left(t_{21}\right)\right)
+ 3\log\!\left(0.869\, t_{21} + 0.034\, t_{21}^2\right)$
\\
\vspace{0.01cm}\\
\hline
\end{tabular}
\caption{Symbolic expressions for the coefficients $\alpha_i$ in the Ansatz~\eqref{eq:phiGinv}, as a function of $t_{12}=t_1/t_2$ or $t_{21}=t_2/t_1$.}
\label{tab:alpha_fits}
\end{table}

As a final step we apply symbolic regression to the numerical data in Fig.~\ref{fig:bicubic_alpha_plots_symmetrised}, using \texttt{PySR}~\cite{PySR} with the building blocks $+$, $\cdot$, $\exp$, $\log$, $\sin$ and $\cos$. The choice of these building blocks is not motivated by any theoretical considerations; rather, it reflects a standard setup used in symbolic-regression algorithms. Internally, \texttt{PySR} employs an evolutionary (genetic) algorithm, in which candidate expressions evolve through successive generations. In our numerical experiment, the algorithm was run for 300 generations on 10 independent populations, each consisting of 200 candidate expressions. The analytic expressions listed in Table~\ref{tab:alpha_fits} were selected from the resulting Pareto front (the set of candidate expressions that are optimal in the sense that no other candidate simultaneously achieves both a lower fitting error and a lower complexity). Inserting these expressions, together with the invariants~\eqref{eq:singlets}, into the Ansatz~\eqref{eq:FS_metric}, \eqref{eq:phiGinv} yields an analytic K\"ahler potential whose average deviation from the numerically learned potential is $2.2\%$.

Finally, to test the accuracy of the proposed method, we proceed by comparing the $\sigma$-loss of the numerical learned metric to that of the analytic one. The $\sigma$-loss is defined as
\begin{equation}
    \sigma_{\rm loss} ~=~ \frac{1}{N_{\rm pts}} \,\left|\left| 1 - \frac{\det g_{\rm pr}}{\kappa \, \Omega \wedge \overline{\Omega}} \right|\right|_{1} \, ,
\end{equation}
where $g_{\rm pr}$ is the predicted Ricci-flat metric, $\Omega$ the holomorphic (3,0) form and $\kappa$ a numerical constant (once the CY volume is fixed). As defined above, it corresponds to the $L_1$-norm deviation of the predicted metric from the solution of the Monge--Amp\`ere equation, averaged over the point sample. 

\begin{figure}[h!]
    \centering
    \raisebox{-0.26cm}{
    \includegraphics[width=0.38\textwidth]{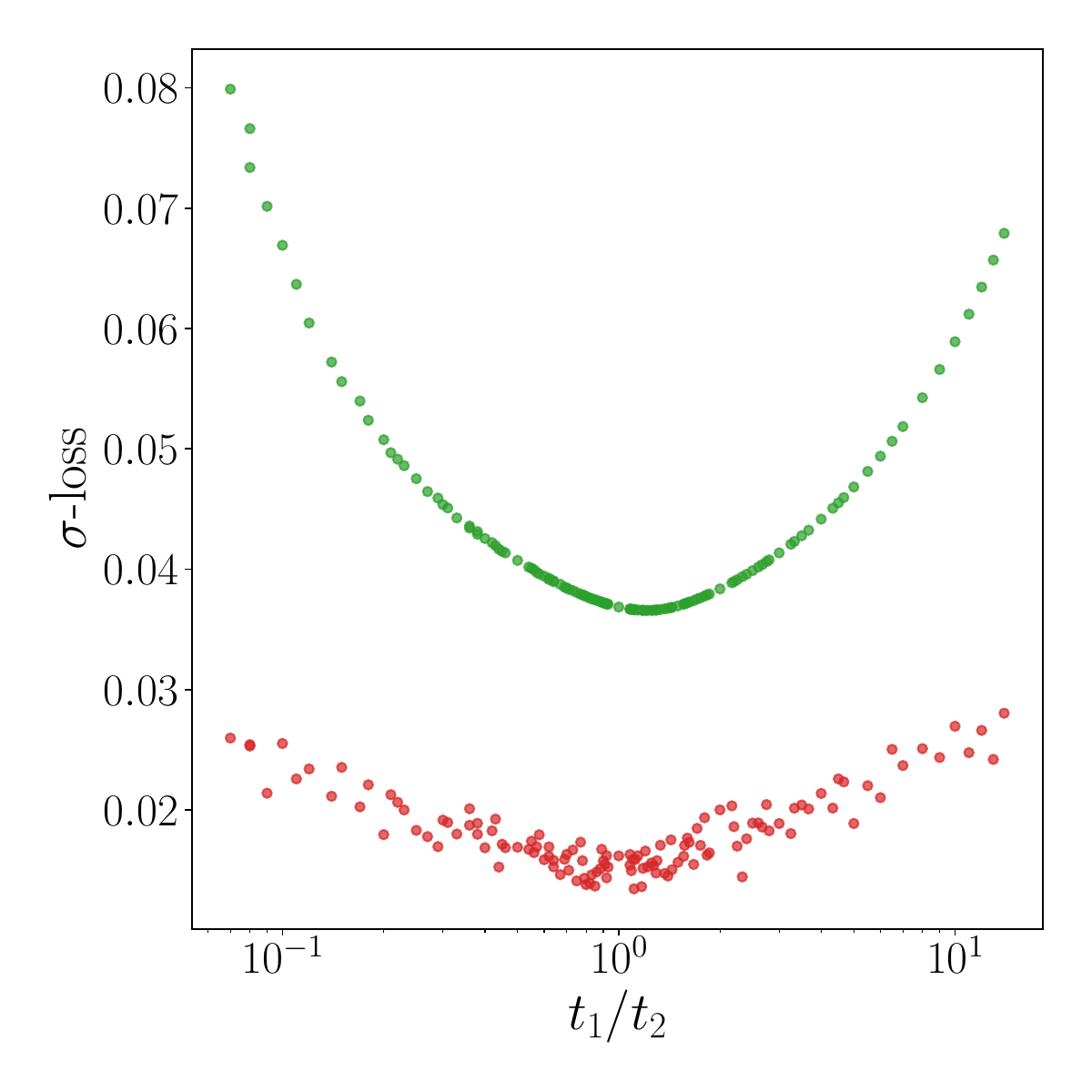}
    }
    \caption{In green, the $\sigma$-loss computed from the analytical expression obtained via symbolic regression, in the case $k_1 = k_2 = 2$. In red, the $\sigma$-loss achieved by the neural network after training. Both quantities are plotted as functions of $t_1/t_2$.}
    \label{fig:bicubic_avg_deviation_sigma_loss}
\end{figure}
Fig.~\ref{fig:bicubic_avg_deviation_sigma_loss} shows how the Ricci-flatness of the numerical learned metric compares with that of the analytic one obtained via symbolic regression.  Given the above definition, we can conclude that although the numerical metric shows lower values, the analytic counterpart can nevertheless be regarded as Ricci-flat to a good approximation. These results indicate that the proposed Ansatz is capable of capturing the moduli dependence of the Ricci-flat metric with good accuracy already at low truncation order.

\section{Ricci-flat metrics on the $(2,4)$ hyper-surface in $\mathbb{P}^1 \times \mathbb{P}^3$ }
\label{sec:2_4}
Our second example is a Calabi--Yau hypersurface of bi-degree $(2,4)$ in the ambient space $\mathcal A=\mathbb P^1\times\mathbb P^3$. We denote homogeneous coordinates by $x=(x_\alpha)_{\alpha=0,1}$ on the $\mathbb P^1$ factor and by $y=(y_a)_{a=0,1,2,3}$ on the $\mathbb P^3$ factor. This manifold has Hodge numbers $h^{1,1}(X)=2$ and $h^{2,1}(X)=86$. As in the bicubic example, we focus on a special point in complex structure moduli space with a comparatively large discrete symmetry group.

In terms of the K\"ahler parameters $t=(t_1,t_2)$, the volume is
\begin{equation}
V(t)=\frac13\,t_2^2(6t_1+t_2)\, ,
\end{equation}
and the K\"ahler cone is specified by $t_1>0$ and $t_2>0$. Although this geometry is still relatively simple, it is intrinsically less symmetric than the bicubic, and it therefore provides a useful test of our method in a less constrained setting.

\subsection{The choice of $(2,4)$ hyper-surface}
We choose a point in moduli space where the manifold has a symmetry $G=\mathbb{Z}_2^{(x)}\times S_4^{(z)}$, with the two factors acting as
\begin{equation}
    \begin{array}{rclclcl}
        \mathbb{Z}_2^{(x)}&:& x_0\leftrightarrow x_1&,& y_a\mapsto y_a\;\;\mbox{for}\;\; a=0,1,2,3\\[2mm]
        S_4^{(z)}&:&x_b\mapsto x_b\;\;\mbox{for}\;\;b=0,1&,&y_a\mapsto y_{\sigma(a)}\;\;\mbox{for}\;\;\sigma\in S_4\;\;\mbox{and}\;\;a=0,1,2,3
    \end{array}\; .
\end{equation}
It turns out that the space of $G$-invariant polynomials of bi-degree $(2,4)$ is $10$-dimensional and spanned by the products $Q_iP_j$, where $i=1,2$ and $j=1,\ldots ,5$ and
\begin{equation*}
\begin{array}{rclcrclcrclcrclc}
Q_1&=&x_0^2+x_1^2&~~~&Q_2&=&x_0x_1&~~~&P_1 &=&\ds \sum_{a} y_a^4&~~~&&&\\[4mm]
   P_2 &=& \ds\sum_{a<b} y_a^2 y_b^2&&P_3 &=&\ds\sum_{a \neq b} y_a^3 y_b&&P_4&=&y_0 y_1 y_2 y_3&&P_5&=&\ds\sum_{\substack{a,b,c\ \mathrm{distinct}\\ b<c}} y_a^2 y_b y_c \, .
\end{array}\;    
\end{equation*}
Within this $10$-dimensional space of $G$-invariant manifolds we choose the following defining polynomial which leads to a smooth Calabi--Yau threefold:
\begin{equation} 
 P = Q_1\,(P_1 - P_3) + Q_2\,P_2~.
 \end{equation} 

\subsection{Invariants}\label{sec:inv24}
The entire symmetry group $G$ defined in the previous sub-section acts trivially on the K\"ahler structure, so, using the notation of Eq.~\eqref{GH} we have $G=N$. Later, we will focus on the line bundle $L={\cal O}_X(2,2)$ with $h^0(X,L)=30$ and an associated matrix $\alpha_{IJ}$ of size $30\times 30$. It turns out there are $30$ $G$-singlets contained in $\Gamma(X,L)\times \Gamma(X,L)^*$ which include the `diagonal' invariants
\begin{align}
\label{eq:singlets_2_v2}
\begin{alignedat}{3}
I_0 \;&=\; \frac{1}{8}\sum_{\alpha,a} |x_\alpha|^4 |y_a|^4 ,
&\qquad
I_1 \;&=\; \frac{1}{12}\sum_{\alpha,a<b} |x_\alpha|^4 |y_a|^2 |y_b|^2 , \\[0.5em]
I_2 \;&=\; \frac{1}{4}\sum_{a} |x_0|^2 |x_1|^2 |y_a|^4  ,
&\qquad
I_3 \;&=\; \frac{1}{6}\sum_{a<b} |x_0|^2 |x_1|^2 |y_a|^2 |y_b|^2  
\end{alignedat}\; .
\end{align}
Unlike in the case of the bi-cubic, the symmetry $G$ does not forbid all off-diagonal invariants. A subset of these is given by
\begin{equation}
\begin{array}{rclcrcl}
I_4  &=&\ds\frac{1}{8}\left(x_0^2 \,\overline{x}_1^2 + \overline{x}_0^2 x_1^2 \right) \,\sum_a |y_a|^4&\;&
I_5  &=&\ds \frac{1}{12}\left(x_0^2 \,\overline{x}_1^2 + \overline{x}_0^2 x_1^2 \right) \,\sum_{a<b} |y_a|^2 \,|y_b|^2\\[8mm]
I_6 &=&\ds \frac{1}{24}|x_0|^2 |x_1|^2\sum_{ a \neq b }  |y_a|^2 \,\left(\overline{y}_b \, y_b \,+\, \text{c.c}  \right)&& 
I_7 &=&\ds \frac{1}{24}\sum_{\alpha \neq \beta}  \left(|x_\alpha|^2 \, \overline{x}_\alpha \, x_\beta  + \text{c.c.} \right) \sum_{a<b}|y_a|^2 |y_b|^2\\[8mm]
I_8  &=&\ds \frac{1}{24} \sum_\alpha |x_\alpha|^2  \,\sum_{ a \neq b }  y_a^2 \, \overline{y}_b^2&&
I_9 &=&\ds \frac{1}{48} \sum_{\alpha \neq \beta}  |x_\alpha|^2  \overline{x}_\alpha  x_\beta   \sum_{a \neq b}|y_a|^2 \,y_a  \overline{y}_b ~+~ \text{c.c}\\[8mm]
I_{10}  &=&\ds \frac{1}{48} \sum_{\alpha \neq \beta} |x_\alpha|^2   \overline{x}_\alpha  x_\beta  \sum_{ \substack{a\neq b,c \\ b<c} } y_a^2 \overline{y}_b  \overline{y}_c ~+~ \text{c.c.}&& 
I_{11} &=&\ds \frac{1}{48}\sum_{\alpha } |x_\alpha|^4 \sum_{ a \neq b }  |y_a|^2 \,\overline{y}_a \, y_b ~+~ \text{c.c}
\end{array}\; .
\label{eq:singlets_3_v2}
\end{equation}
There are $18$ further off-diagonal invariants not listed here which will turn out to be sub-dominant.

\subsection{Numerical results}

We follow the same procedure as for the bicubic and choose 127 different numerical values for the ratio $t_1/t_2$ of the two \K moduli, as in Eq.~\eqref{t12val}. Using \texttt{cymetric}, we train a fully connected neural network with the same architecture as in the previous case (width 64, depth 3, GeLU activation) and the same `adiabatic' training strategy: we begin with a full training cycle for the first value of $t_{12}=t_1/t_2$ and then proceed to nearest neighbours in $t_{12}$, using the trained parameter of the previous run as initialisation for the next.

\begin{figure}[h!]
    \centering    \includegraphics[width=0.415\textwidth]{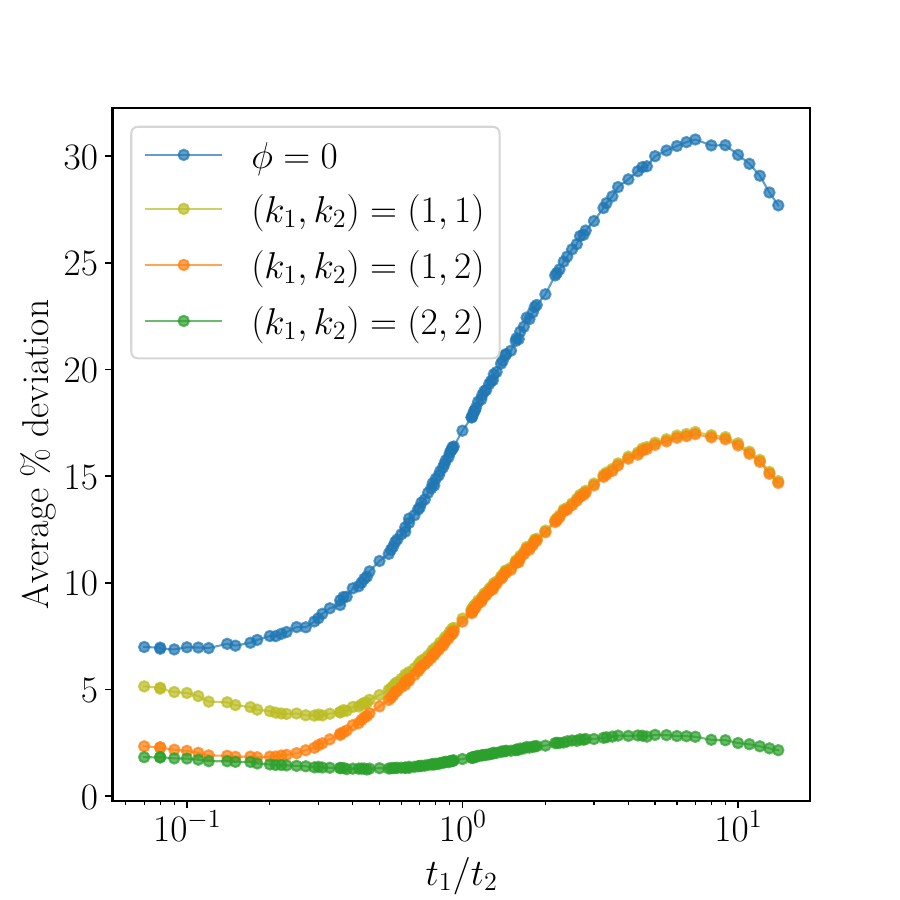}
    \caption{Average relative deviation $\langle |K_{\rm NN}-K|/K_{\rm NN}\rangle$ between the analytic \K potential $K$ and the numerically learned potential $K_{\rm NN}$, shown as a function of $t_1/t_2$ for different truncations of the Ansatz. The blue points correspond to the pure Fubini--Study potential ($\phi=0$), while the remaining points correspond to increasing values of $(k_1,k_2)$ as indicated in the legend. Averaged over all sampled values of $t_1/t_2$, the deviations are $18\%$, $9\%$, $8.5\%$, and $1.9\%$, in the order given in the legend.
}
    \label{fig:2,4_avg_deviation}
\end{figure}

Fitting the numerical value of the \K potential obtained from the neural network with the Ansatz in Eq.~\eqref{eq:phi}, we get the results in Fig.~\ref{fig:2,4_avg_deviation}. The plot shows the error of the fit for $\phi = 0$ (blue curve), which corresponds the Fubini-Study metric, as well the errors for $(k_1,k_2)=(1,1),(1,2)$ and $(2,2)$. As expected, the most accurate fit is obtained for the highest choice of $k_1$ and $k_2$. 
Moreover, we observe that the curve corresponding to $(k_1,k_2)=(1,2)$ displays deviations comparable to those of the $(2,2)$ Ansatz in the regime where the $x$-direction is strongly squeezed, $t_1/t_2 \sim 10^{-1}$. By contrast, as the ratio $t_1/t_2$ increases, the same $(1,2)$ curve smoothly approaches and eventually merges with the $(1,1)$ case. This behaviour can be understood geometrically. For small $t_1/t_2$, the \K class suppresses variations along the $\mathbb{P}^1$ directions, so the dominant contribution to the deviation originates from how accurately the Ansatz captures the geometry along the $\mathbb{P}^3$ factor. In this regime, the error is therefore controlled primarily by the degree $k_2$, explaining why the $(1,2)$ and $(2,2)$ ansätze perform similarly. Conversely, when $t_1/t_2$ becomes large, the roles of the two factors are effectively exchanged: the geometry is now stretched along $\mathbb{P}^1$, and the deviation is governed almost entirely by the choice of $k_1$, leading to the observed agreement between the $(1,2)$ and $(1,1)$ curves.

\begin{figure}[h!]
    \centering
    \includegraphics[width=0.32\textwidth]{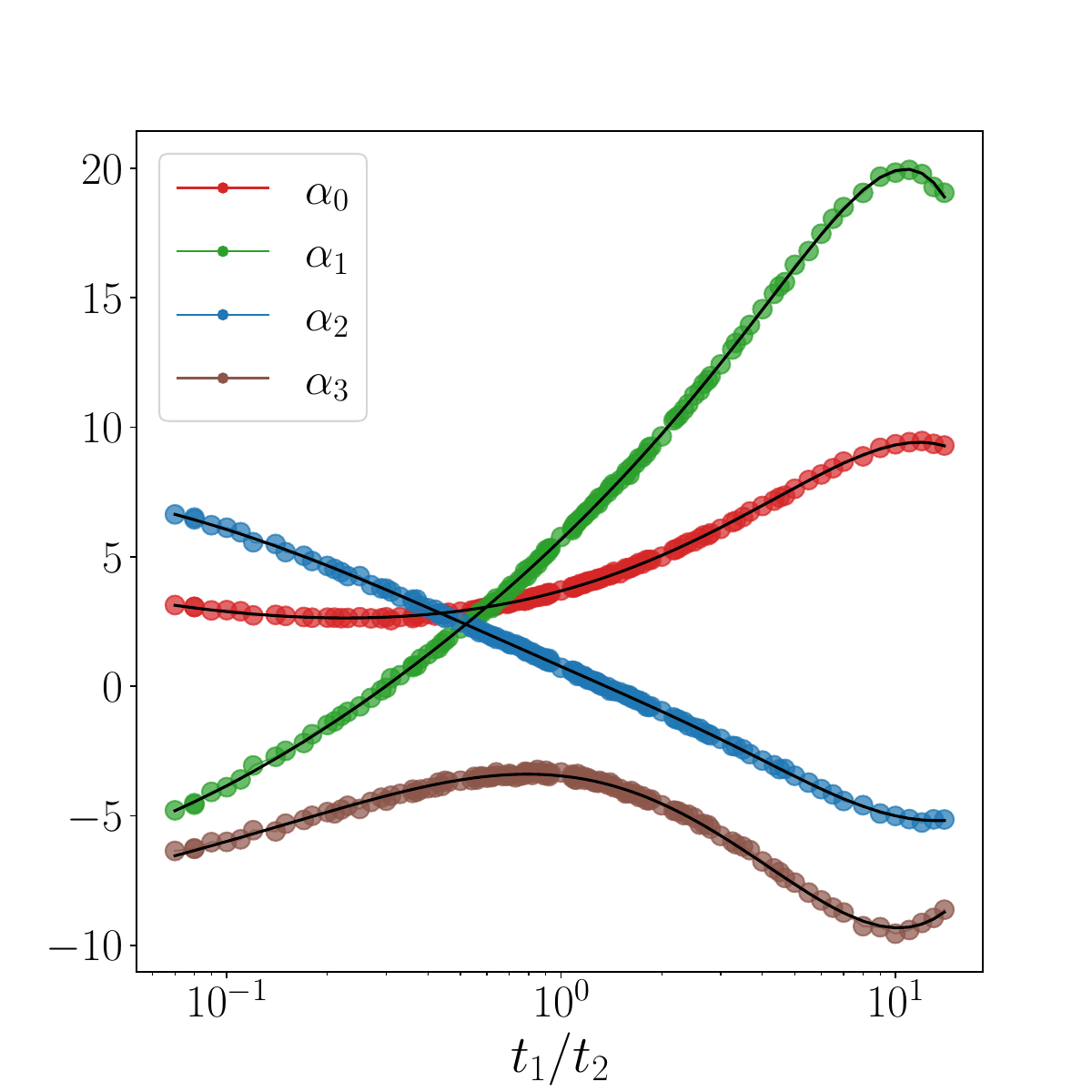}
    \includegraphics[width=0.32\textwidth]{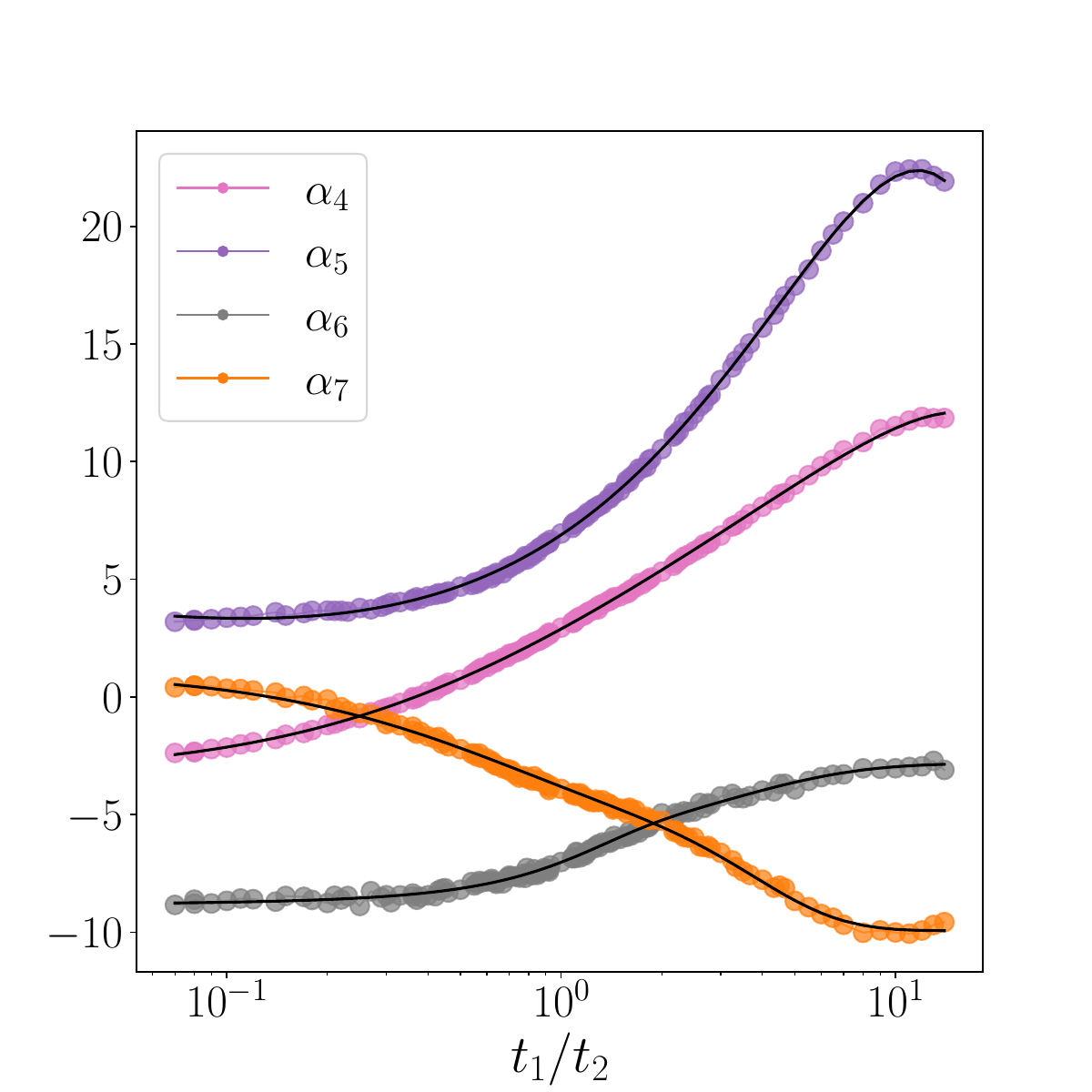}
    \includegraphics[width=0.32\textwidth]{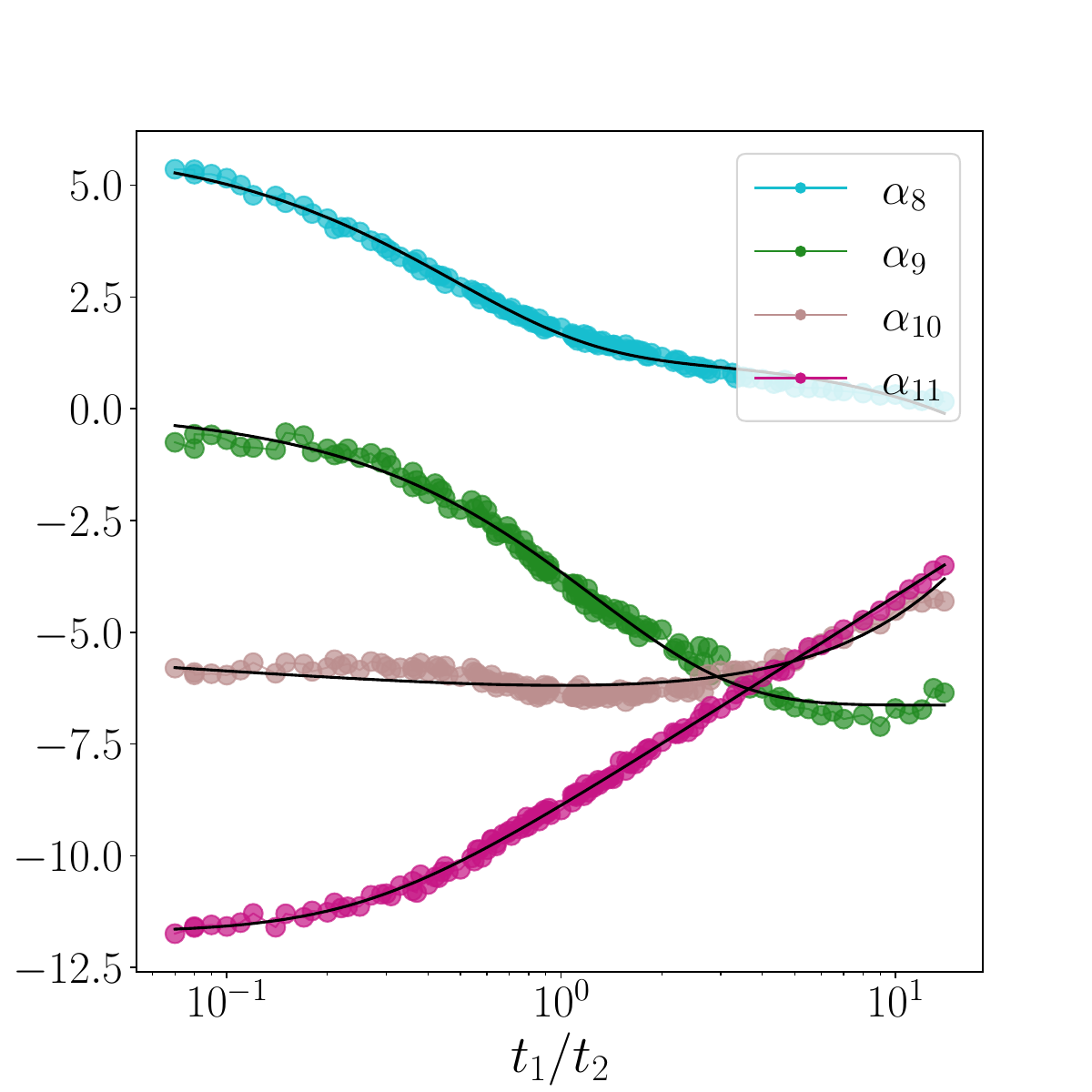}
    \caption{Monomial coefficients $\alpha_i$ as functions of $t_1/t_2$ in the $(k_1,k_2)=(2,2)$ truncation. The left panel displays the diagonal coefficients, while the other two panels present the leading off-diagonal ones. The black lines corresponds to the analytic expression obtained using symbolic regression.}
    \label{fig:2_4_alpha_plots_2}
\end{figure}

As for the bi-cubic, we numerically checked that the function $\phi$ is $G$-invariant. Therefore, focusing on the most accurate case of $(k_1,k_2)=(2,2)$ we can work with the Ansatz~\eqref{eq:phiGinv} and the invariants $I_i$, where $i=0,\ldots ,11$, listed in the previous sub-section. Fig.~\ref{fig:2_4_alpha_plots_2} shows the fitted values of the associated coefficients $\alpha_i$ in Eq.~\eqref{eq:phiGinv}.

\begin{figure}[h!]
    \centering
\includegraphics[width=0.4\textwidth]{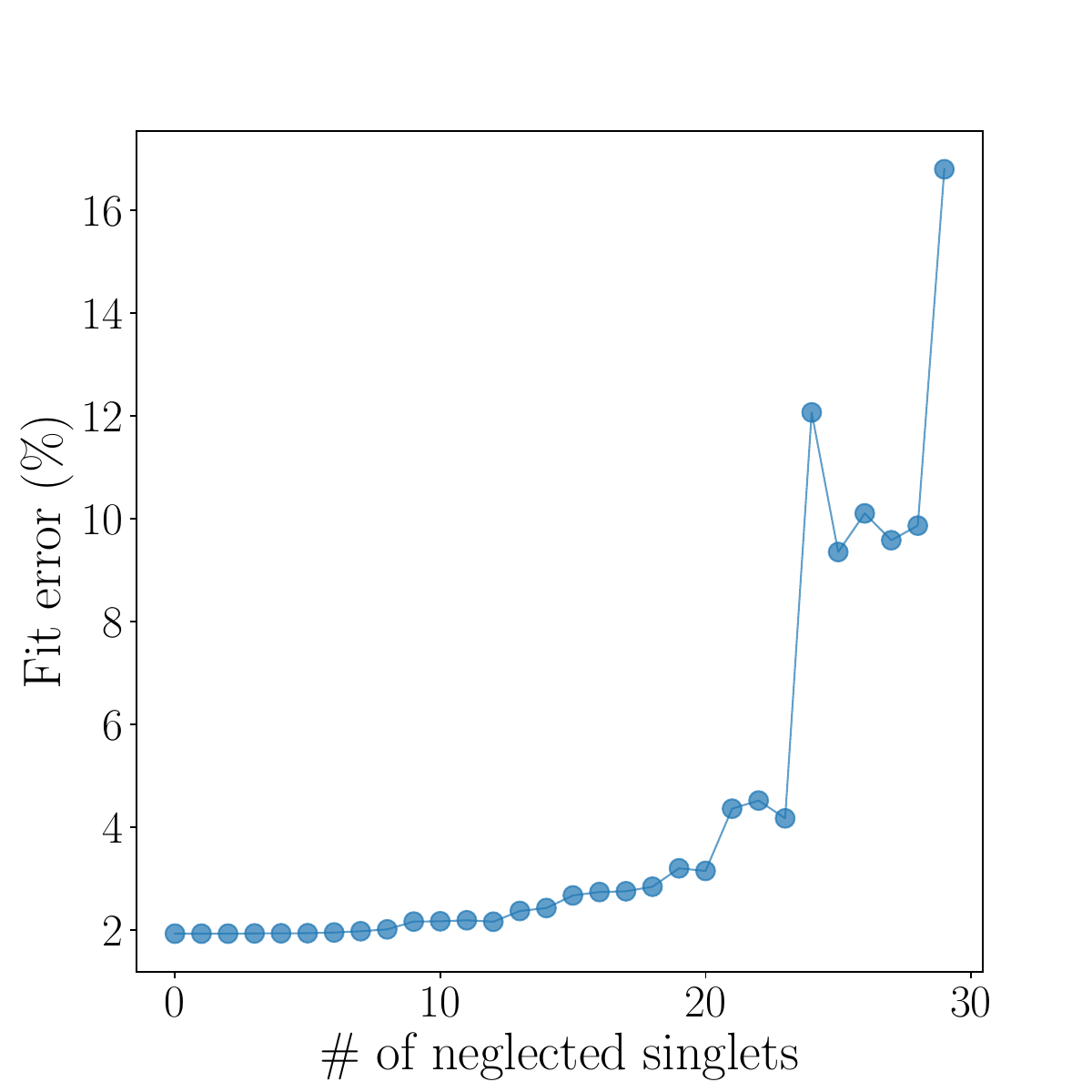}
    \caption{Fit error as a function of the number of neglected singlets in the analytic \K potential. Singlets are ordered according to the size of their  contribution, quantified by $\max_{t_{12}} |\alpha_i|$, from smallest to largest (left to right over the $x$-axis). }    \label{fig:n_of_neglected_singlets_vs_error}
\end{figure}

Fig.~\ref{fig:n_of_neglected_singlets_vs_error} shows how the fit error increases with the number of singlet contributions that are ignored in the Ansatz~\eqref{eq:phiGinv} (meaning their associated coefficient are set to zero `by hand'). The plot shows that up to approximately twenty singlets can be neglected without a significant deterioration in accuracy. Hence, retaining only the $12$ singlets listed in the previous sub-section constitutes a good approximation which achieves an accuracy of $2.8$\%. The neglected singlets are associated with corresponding $\alpha$ coefficients that are suppressed by one to two orders of magnitude compared to those of the retained singlets and are approximately constant throughout the sampled region of moduli space.

For these $12$ coefficients we carry out a symbolic regression, using {\sf PySR} with building blocks $+$, $\cdot$, $\exp$, $\log$, $\sin$ and $\cos$, using the numerical data shown in Fig.~\ref{fig:2_4_alpha_plots_2}.
The algorithm evolves 10 independent populations of 200 individuals over 300 generations. The analytical expressions reported in Table~\ref{tab:24_alpha_fits} were selected from the corresponding Pareto fronts. Inserting these expressions together with the invariants in Section~\ref{sec:inv24} into the Ansatz~\eqref{eq:phiGinv} gives the approximate analytic form of the Ricci-flat K\"ahler potential.

\begin{figure}[h!]
    \centering
    \raisebox{-0.26cm}{
    \includegraphics[width=0.4\textwidth]{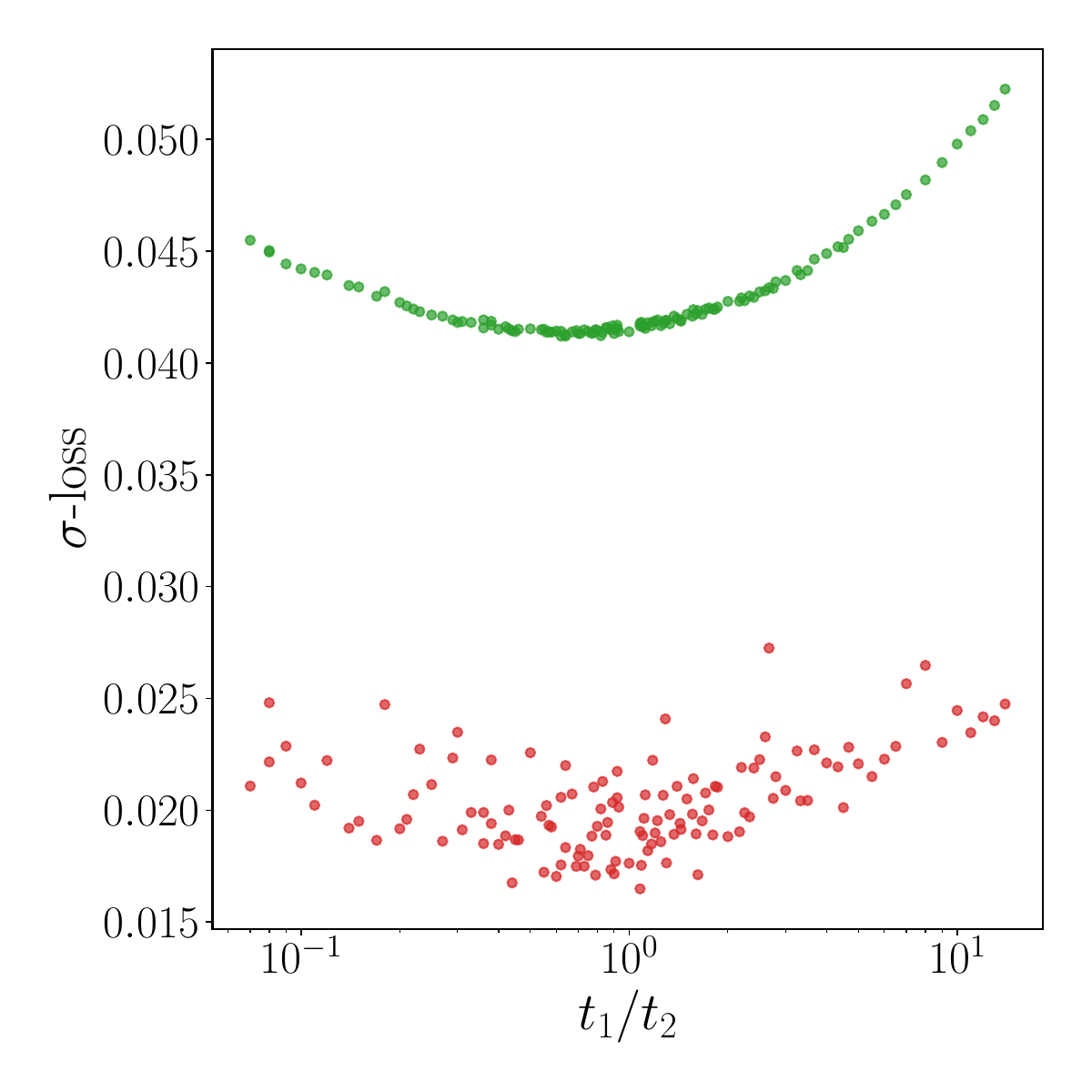}
    }
    \caption{In green, we show the $\sigma$-loss computed from the analytic expression obtained via symbolic regression for $k_1 = k_2 = 2$. In red, we show the $\sigma$-loss achieved by the trained neural network. Both quantities are plotted as functions of $t_1/t_2$.
}
\label{fig:2,4_avg_deviation_sigma_loss}
\end{figure}
As in the case of the bicubic, we computed the $\sigma$-loss of the analytic metric and compared to that of the numerical one. We plotted these quantities in Fig.~\ref{fig:2,4_avg_deviation_sigma_loss}. As in the bicubic case, the proposed Ansatz yields a moduli-dependent metric that is approximately Ricci-flat even at relatively low truncation order. 
\begin{table}[!ht]
\centering
\begin{tabular}{c l}
\hline
Coefficient & Fit expression \\ 
\hline
\vspace{0.01cm}\\

$\alpha_0$ &
$1.50 + 2.18\,t_{12}
- 0.65\,t_{12}\log(t_{12})
+ 0.19\,\log^2(t_{12})$
\\[6pt]

$\alpha_1$ &
$4.07 + 1.44\,t_{12}
- 0.086\,t_{12}^2
+ 4.35\,\log\!\bigl(0.057 + t_{12}\bigr)$
\\[6pt]

$\alpha_2$ &
$0.28\,t_{12}
+ 1.12\,t_{12}e^{-0.40\,t_{12}}
- 3.47\,\log\!\bigl(0.0818 + t_{12}\bigr)$
\\[6pt]

$\alpha_3$ &
$-5.25 + 1.78\,t_{12}
- 4.00\,\log(t_{12})
\,\log\!\bigl(0.805 + t_{12}\bigr)$
\\[6pt]

$\alpha_4$ &
$\bigl(4.52 - 0.0026\,t_{12}^2\bigr)
\,\log\!\bigl(0.48 + 1.41\,t_{12}\bigr)$
\\[6pt]

$\alpha_5$ &
$-0.45 + 7.33\,t_{12}
- (1.12\,
+ 2.09\,t_{12})\, \log(t_{12})$
\\[6pt]

$\alpha_6$ &
$-2.83
- 4.76\,e^{-0.36\,t_{12}}
- 1.23\,e^{\,0.65\,t_{12}-t_{12}^2}$
\\[6pt]

$\alpha_7$ &
$-9.94
+ (11\,
+ 3.44\,t_{12}^2) e^{-0.86\,t_{12}}$
\\[6pt]

$\alpha_8$ &
$1.2
+ 24\,e^{-1.62 - 2.13\,t_{12}}
- 0.093\,t_{12}$
\\[6pt]

$\alpha_9$ &
$3.31(e^{-0.8\,t_{12}}-1)$
\\[6pt]

$\alpha_{10}$ &
$-6.42
+ 0.23\,(t_{12}
- \log(t_{12}))$
\\[6pt]

$\alpha_{11}$ &
$-8.93
+ 1.03\,\log\!\bigl(0.068 + t_{12}^2\bigr)$
\\[6pt]

$\alpha_{12,...,29}$ &
$\stackrel{!}{=} 0$
\\[6pt]
\hline
\end{tabular}
\caption{Analytic expressions, obtained via symbolic regression, for coefficients $\alpha_i(t_{12})$, $i=0,\dots,11$, as functions of $t_{12} \equiv t_1/t_2$. The remaining coefficients are sub-leading and therefore manually set to zero.}
\label{tab:24_alpha_fits}
\end{table}

\section{Conclusions}
\label{sec:conclusions}

In this work, we have developed a framework for constructing approximate analytic expressions for Ricci-flat metrics on Calabi--Yau three-folds with explicit dependence on the \K moduli. Our strategy combines numerical data obtained from machine-learning techniques with an analytic Ansatz for the \K potential built from sections of ample line bundles. Neural networks are first used to learn the \K potential across a sampling of the \K moduli space; the resulting data are then fitted with analytic expressions whose coefficients are promoted to explicit functions of the \K parameters. In this way, we obtain closed-form analytic approximations to the Ricci-flat \K potential that retain its non-trivial moduli dependence while remaining fully symbolic. Computing the $\sigma$-loss of the associated analytic metric and comparing it with the numerical metric obtained from the neural network, we find that the analytic metric remains Ricci-flat to a good approximation. To our knowledge, these examples of approximate Ricci-flat metrics with explicit K\"ahler moduli dependence are the first of their kind.

A particularly striking outcome of our analysis is that relatively small values of the line-bundle integers used in the Ansatz already yield analytic expressions that agree with the numerically learned \K potentials at percentage level in the sampled region of the moduli space. The agreement extends beyond the level of the K\"ahler potential to the metric itself: in both examples analysed, the analytic metric achieves sufficiently small $\sigma$-loss values to be regarded as approximately Ricci-flat. This is far from obvious a priori: the Ansatz corresponds to a highly truncated finite-dimensional space, yet it captures the dominant geometric variations with remarkable accuracy. 

We have applied this strategy to two Calabi--Yau three-folds with $h^{1,1}(X)=2$, namely a bi-cubic hypersurface in $\mathbb{P}^2 \times \mathbb{P}^2$ and a bi-degree $(2,4)$ hypersurface in $\mathbb{P}^1 \times \mathbb{P}^3$. In both case, we have chosen a single point in complex structure moduli space such that the associated manifold has a large discrete symmetry. As a consistency check of both the numerical fit and the truncation scheme, we employed a symmetry-agnostic Ansatz and verified numerically that the resulting fit reproduces the same symmetry. These symmetries considerably simplify the structure of our analytic Ansatz as basis functions are forced to combine into invariant combinations. For the bi-cubic manifold, this leads to a compact and highly accurate analytic expression. For the second example, despite a smaller symmetry group and a more complicated Ansatz, most invariants turn out to have sub-leading contributions and can be neglected with only a marginal increase in error. 

There are several natural directions for further development. First, while the present work focuses on K\"ahler moduli dependence, extending the framework to incorporate complex-structure moduli would be highly desirable. Achieving simultaneous analytic control over both sectors would represent a significant step towards fully moduli-dependent analytic approximations of Calabi--Yau metrics and would substantially broaden the scope of phenomenological applications. Second, one could attempt to bypass the intermediate neural-network stage and use symbolic regression directly, incorporating the Ricci-flatness condition into the loss function from the outset. In such an approach, one would aim to learn analytic expressions for the \K potential by directly minimising a functional encoding the Monge--Amp\`ere equation. Although computationally demanding~\cite{Mirjanic:2024gek}, especially in higher-dimensional moduli spaces, improved implementations and parallelisation strategies may render this approach viable. An important avenue for future work is to determine how large the line bundle degree must be taken in order for the analytically reconstructed metric to achieve a level of Ricci-flatness comparable to that of the numerical metric. While the results presented here show that the analytic metric remains Ricci-flat to a good approximation, a noticeable gap persists relative to the neural-network solution. It remains to be seen whether this discrepancy can be reduced, or even eliminated, by constructing Ansatzes associated with higher-degree line bundles. Equally important is to assess the extent to which the current analytic approximations already suffice for the computation of metric-dependent physical quantities.

\section*{Acknowledgements}

AC and LAN are supported by the Royal Society grant DHF/R1/231142. AL acknowledges support from the STFC consolidated grant ST/X000761/1. We are grateful to Jonathan Patterson for his assistance with the Oxford
Theoretical Physics computing cluster \texttt{Hydra}.
We would like to thank Steve Abel and Miguel Crispim Rom\~ao for valuable discussions and for their collaboration during an earlier stage of this project. 
We are grateful to the anonymous referee for pointing out the argument included in footnote 1, which establishes that holomorphic automorphisms of the CY manifold are respected by the Ricci-flat metric.
Finally, we thank the International Centre for Mathematical Sciences (ICMS), Edinburgh, for hospitality during the final stages of writing this paper.
\printbibliography
\end{document}